\documentstyle[11pt,aaspp4]{article}
 
\begin{document}
 
\title{Observations of the SW Sextantis Star UU Aquarii}

\author{D. W. Hoard,\footnotemark[1] M. D. Still,\footnotemark[2] Paula 
Szkody,\footnotemark[1] Robert Connon Smith,\footnotemark[3] and D. A. H. 
Buckley\footnotemark[4]}
\footnotetext[1]{Department of Astronomy, University of Washington, Box 351580, 
Seattle, WA 98195-1580, USA; email: hoard@astro.washington.edu, 
szkody@astro.washington.edu}
\footnotetext[2]{Physics \& Astronomy, University of St.\ Andrews, North Haugh, 
St.\ Andrews, Fife KY16 9SS, UK; email: martin.still@st-and.ac.uk}
\footnotetext[3]{Astronomy Centre, University of Sussex, Brighton BN1 9QJ, UK; 
email: rcs@star.maps.susx.ac.uk}
\footnotetext[4]{South African Astronomical Observatory, PO Box 9, Observatory 
7935, Cape Town, South Africa; email: dibnob@saao.ac.za}

\slugcomment{accepted for publication in MNRAS}
 
\begin{abstract}
We present 14 nights of medium resolution (1-2\AA) spectroscopy of the 
eclipsing cataclysmic variable UU Aquarii obtained during a high accretion 
state in 1995 August--October.  UU Aqr appears to be an SW Sextantis star, as 
noted by \markcite{BSH96}Baptista, Steiner, \& Horne (1996), and we discuss its 
spectroscopic behavior in the context of the SW Sex phenomenon.  Emission line 
equivalent width curves, Doppler tomography, and line profile simulation 
provide evidence for the presence of a bright spot at the impact site of the 
accretion stream with the edge of the disk, and a non-axisymmetric, vertically- 
and azimuthally-extended absorbing structure in the disk.  The absorption has 
maximum depth in the emission lines around orbital phase 0.8, but is present 
from $\phi\approx0.4$ to $\phi\approx0.95$.  An origin is explored for this 
absorbing structure (as well as the other spectroscopic behavior of UU Aqr) in 
terms of the explosive impact of the accretion stream with the disk.  
\end{abstract}


\section{Introduction}
\label{s-intro}

Cataclysmic variables (CVs) are short period (P$_{orb}$ typically $\lesssim 1$ 
d), semi-detached binary systems in which matter is transferred through the 
$L_{1}$ point from a Roche-lobe-filling, lower main sequence dwarf (the 
secondary star), into an accretion disk around the white dwarf (WD) primary 
star.  In most nonmagnetic CVs, energy output from the disk dominates the 
optical--X-ray luminosity; hence, the properties of the disk (which are 
primarily governed by the mass transfer rate $\dot{M}$ in the case of an 
optically thick disk) determine most of the observed properties of the entire 
system.  UU Aquarii ($V\approx13.5$) was identified as a variable star early in 
this century (\markcite{Beljawsky26}Beljawsky 1926).  As observations of UU Aqr 
have been accumulated, its classification has evolved from a long-period 
semi-regular variable (\markcite{Payne52}Payne-Gaposchkin 1952), to a low 
$\dot{M}$, optically thin disk, dwarf nova type CV (\markcite{Volkov86}Volkov, 
Shugarov, \& Seregina 1986), to a high $\dot{M}$, optically thick disk, 
novalike CV (\markcite{DS91}Diaz \& Steiner 1991, henceforth DS91; 
\markcite{BSC94}Baptista, Steiner, \& Cieslinski 1994, henceforth BSC94).

\markcite{Volkov86}Volkov et al.\ (1986) and \markcite{Goldader89}Goldader \& 
Garnavich (1989) observed deep ($\approx1.5$ mag) eclipses in UU Aqr, and 
established an orbital period of P$_{orb}=0.163579089(61)$ d 
($\approx3.93$ h).  This places UU Aqr in the 3--4 hr period range just above 
the 2--3 hr CV period gap (thought to be due to the transition from orbital 
angular momentum loss primarily via magnetic braking for P$>3$ hr to loss due 
to gravitational radiation for P$<2$ hr; \markcite{King88}King 1988).  Among 
novalike CVs, the 3--4 hr period range is occupied almost solely by the group 
of systems known as the {\em SW Sextantis stars} (e.g., 
\markcite{Thorst91}Thorstensen et al.\ 1991), which are characterized by high 
inclination, single-peaked optical emission lines (in contrast to the 
double-peaked lines expected in high inclination disk systems), 
orbital-phase-dependent absorption in the Balmer lines, and a phase offset in 
the emission line radial velocity curves (implying non-uniform emission from 
the disk).  As a result of their multiwavelength eclipse mapping study, 
\markcite{BSH96}Baptista, Steiner, \& Horne (1996; henceforth BSH96) conclude 
that UU Aqr is most likely an SW Sex star (a conclusion that we share based on 
the results presented here -- see \S\ref{s-disc}).  

\markcite{BSC94}BSC94 obtained photometric data of UU Aqr on 28 nights during 
1988--1992.  They discovered that this system, like many other novalike CVs 
(\markcite{Dhillon96}Dhillon 1996), exhibits occasional transitions from high 
to low brightness states.  The amplitude of the state change in UU Aqr, 
however, is only $\approx0.2$--$0.3$ mag, which is small compared to other 
novalike CVs (e.g., BH Lyncis: $\Delta$mag$\approx1.5$--$2.0$, 
\markcite{Hoard97}Hoard \& Szkody 1997; DW Ursae Majoris: 
$\Delta$mag$\approx2$--$3$, \markcite{Dhillon94}Dhillon, Jones, \& Marsh 
1994).  The shape of the eclipse profile in UU Aqr does change significantly 
from the high to low states, and \markcite{BSC94}BSC94 concluded that the main 
difference between the two states is the presence of a bright spot on the outer 
edge of the disk in the high state.

Both \markcite{Haefner89}Haefner (1989) and \markcite{DS91}DS91 determined 
radial velocity curves for the H$\alpha$ line in UU Aqr; both noted the 
strongly asymmetric, phase-dependent profile of the line.  
\markcite{Haefner89}Haefner (1989) found a velocity semi-amplitude of 160 km 
s$^{-1}$ from the wings of the H$\alpha$ line, while \markcite{DS91}DS91 
obtained 120 km s$^{-1}$ using the same method.  \markcite{DS91}DS91 determined 
system parameters of $q = $ (M$_{2}$/M$_{wd} =$) $0.4$ with M$_{wd} = 
0.9$M$_{\odot}$.  More recently, the detailed eclipse-mapping study of UU Aqr 
by \markcite{BSC94}BSC94 and \markcite{BSH96}BSH96 provided a photometric model 
that requires $K_{wd} \approx 84$ km s$^{-1}$, substantially smaller than 
estimates made from the H$\alpha$ emission line.  (This is not surprising as 
low excitation emission lines, which form primarily in the outer parts of the 
disk, cannot be expected to accurately map the orbital motion of the WD; e.g., 
\markcite{Still95}Still, Dhillon, \& Jones 1995, \markcite{Shafter88}Shafter, 
Hessman, \& Zhang 1988.)  \markcite{BSC94}BSC94 derive system parameters of 
$q=0.30$ with M$_{wd}=0.67$M$_{\odot}$ and an inclination $i=78^{\circ}$.

We present here the results of an extensive spectroscopic monitoring campaign 
on UU Aqr.  Previously published time-resolved spectra (\markcite{DS91}DS91) 
were obtained during a low state, while our spectra explore the high state of 
the system (see \S\ref{s-EW}).  In addition, the spectra of DS91 covered only 
200\AA\ around H$\alpha$, whereas our spectra span H$\beta$ to H$\gamma$ on all 
nights, with additional coverage from H$\alpha$ to \ion{He}{1} $\lambda5876$ on 
one night.  Our results include a line profile simulation that reproduces the 
gross characteristics of the orbital-phase-resolved emission line behavior of 
UU Aqr in terms of several discrete emission and absorption components 
(\S\ref{s-linesim}), Doppler tomography resolved over half an orbit of the CV 
(\S\ref{s-halfDT}), and a qualitative model for the system (\S\ref{s-model}).

\section{Observations}

We observed UU Aqr using the 1.9m Radcliffe Telescope at the South African 
Astronomical Observatory (SAAO) on 1995 August 1--8 UT, 1995 August 18--20 UT, 
and 1995 September 19--26 UT.  The observations were made using the two-channel 
Reticon Photon Counting System (RPCS; \markcite{Menzies96}Menzies \& Glass 
1996) with a 1200 line mm$^{-1}$ grating (giving a resolution of 1.8\AA).  
Exposures were 90 s with a total time per spectrum of $\approx97$ s; a total of 
1492 spectra were obtained.  Sky was recorded in one channel of the RPCS and 
sky+object in the other channel.  Following normalization via bias-subtraction 
and division by a tungsten lamp flat-field, the spectra were extracted by 
simply subtracting the sky channel from the sky+object channel.   
The spectra were wavelength-calibrated by comparison to spectra of a Cu-Ar arc 
lamp standard; the usable wavelength range was 4250--5000\AA.  They were 
flux-calibrated via observations of the standard stars L1788 and L7379 
(\markcite{Hamuy94}Hamuy et al.\ 1994).  In order to increase the 
signal-to-noise, the reduced RPCS spectra were averaged into phase bins of 
width $\Delta\phi=0.04$ in blocks of 2 consecutive nights (see 
Table~\ref{t-log}), using the photometric ephemeris of  \markcite{BSC94}BSC94 
(we will refer to these as the ``binned RPCS data sets'').  We also constructed 
sets of phase-binned ($\Delta\phi=0.04$) spectra using $\approx1450$ of the 
total set of RPCS spectra (some of the RPCS spectra were rejected because they 
were extremely weak, possibly due to slit losses; we will refer to this as the 
``combined RPCS data set'').

We also obtained spectra of UU Aqr on 1995 October 12--13 UT using the Double 
Imaging Spectrograph (DIS) on the Apache Point Observatory 3.5m telescope 
(e.g., see \markcite{Hoard97}Hoard \& Szkody 1997). 
Exposure times were 300 s with a total time of $\approx381$ s per spectrum; a 
total of 65 spectra were obtained.  The ``blue'' side of DIS covered a 
wavelength range of 4200--5000\AA\ and the ``red'' side covered 5800-6800\AA, 
with an overall resolution of $\approx1.5$--$2$\AA.  
The spectra were extracted and calibrated using standard IRAF 
routines \markcite{Massey92}(Massey, Valdes, \& Barnes 1992); they were 
wavelength-calibrated via comparison with a He-Ne arc lamp standard.  
Fine corrections to the wavelength calibration were made by examining the 
positions of night sky lines in the spectra.  Both the blue and red 
spectra were flux-calibrated using spectra of the standard stars 
Feige~110 and BD +28$^{\circ}$ 4211 (\markcite{Massey88}Massey et al.\ 1988).

The spectroscopic observations of UU Aqr are summarized in Table~\ref{t-log}.
\begin{table}
\dummytable\label{t-log}
\end{table}

\section{Results}
\subsection{The Average Spectrum}
\label{s-avsp}

The average spectrum of UU Aqr, outside orbital phases 0.9 to 0.1 (the eclipse) 
and uncorrected for any orbital motion, is shown in Figure~\ref{f-avsp}.  
Each blue spectrum from the binned RPCS data sets listed in Table~\ref{t-log} 
is plotted in the figure.  
The spectrum shows prominent Balmer emission lines 
and several weaker \ion{He}{1} emission lines ($\lambda4471$, $\lambda4921$, 
$\lambda5015$, $\lambda5876$, $\lambda6678$).  The \ion{He}{2} $\lambda4686$ 
emission line is weak and blended with a \ion{C}{3}/\ion{N}{3} emission complex 
of similar strength.  All of the emission lines are single-peaked (this is true 
for the individual spectra as well -- for example, see Fig.~\ref{f-simprof} -- 
so this is not just an effect of smearing of the line profiles in the average 
spectrum).  The blue spectrum is quite constant over the $\approx3$ month span 
of the observations.  The continuum from 4200\AA\ to 6800\AA\ follows a power 
law, $F_{\lambda}\propto\lambda^{-\alpha}$ with $\alpha\approx1.7$.  This is 
somewhat smaller than the power law index $\alpha=2.3$ expected from both 
theoretical calculations in which all disk elements radiate as blackbodies 
(\markcite{Pringle81}Pringle 1981) and observations of dwarf novae at outburst 
(\markcite{Szkody85}Szkody 1985).  A relatively flat spectral energy 
distribution is considered typical of novalike CVs (\markcite{Oke82}Oke \& Wade 
1982), and may be related to departures of the disk structure from simple 
theoretical models and/or self-eclipse of the inner disk in high inclination 
systems (e.g., \markcite{Rutten92}Rutten, van Paradijs, \& Tinbergen 1992).

\subsection{Radial Velocities}
\label{s-rvel}

Radial velocities were measured from the wings of the Balmer emission lines in 
all of the binned RPCS data sets, and from the Balmer and helium lines in each 
of the unbinned DIS spectra, using the double Gaussian fitting technique 
described by \markcite{Shafter86}Shafter, Szkody, \& Thorstensen (1986) and 
\markcite{Schneider80}Schneider \& Young (1980).  
Before applying this technique to each line, a section of the spectrum around 
the wavelength of the line center ($\pm3000$ km s$^{-1}$) was converted to a 
uniform (i.e., linear) velocity scale with a step-size of the local velocity 
dispersion, 
\begin{equation}
(c~\Delta\lambda)/ \lambda_{0}~{\rm km~s}^{-1}, 
\end{equation}
where $c$ is the speed of light (km s$^{-1}$), $\lambda_{o}$ is the wavelength 
of the line center (\AA), and $\Delta\lambda$ is the wavelength dispersion 
($\approx$0.8\AA\ for the RPCS spectra, $\approx$1.3\AA\ for the red DIS 
spectra, and $\approx$1.6\AA\ for the blue DIS spectra).  
For the Balmer and \ion{He}{1} $\lambda5876$ lines, the FWHM of the Gaussians 
was set at twice the local velocity resolution of the spectra for each line 
(FWHM$\approx$ 160 km s$^{-1}$ at H$\beta$ to 180  km s$^{-1}$ at H$\gamma$ in 
the RPCS spectra; FWHM$\approx$ 120 km s$^{-1}$ at H$\alpha$ to 230 km s$^{-1}$ 
at H$\gamma$ in the DIS spectra).  The Gaussian separation $a$ was varied in 
steps of 100 km s$^{-1}$ from 500 km s$^{-1}$ to 2500 km s$^{-1}$.  
In order to obtain usable velocities from the weaker \ion{He}{1} $\lambda4471$ 
and \ion{He}{2} $\lambda4686$ lines (in the DIS spectra), the FWHM of the 
Gaussians was increased to 400 km s$^{-1}$.

The equation
\begin{equation}
V(\phi , a) = \gamma (a) - K(a) \sin [2\pi (\phi - \phi_{0}(a))]
\end{equation}
was fitted to the velocity data obtained from the optimum value of the Gaussian 
separation for each line.  Because we are comparing spectra obtained at 
different times, the velocities from each data set were shifted by the 
appropriate heliocentric correction prior to fitting.  
Representative samples of the diagnostic diagrams (from the DIS data) used to 
determine the optimum Gaussian separation are shown in Figure~\ref{f-diags}.  
For the Balmer lines, the diagrams are well-behaved and we chose $a=1700$ km 
s$^{-1}$.  The diagrams for the helium lines, on the other hand, are rather 
more uncertain.  All three He lines display a trough in the fractional 
uncertainty of the velocity semi-amplitude, $\sigma_{K}/K$, between $a=1100$ km 
s$^{-1}$ and $a=1800$ km s$^{-1}$, indicating that this range of Gaussian 
separations best samples the line wings away from the highly variable line 
cores but before the adjacent continuum contaminates the wings significantly.  
However, the values of the system parameters (especially $K$) still fluctuate 
significantly over this range of $a$.  The \ion{He}{1} $\lambda5876$ line is 
the best-behaved of the He lines, and we have used $a=1700$ km s$^{-1}$ for 
this line also, based on the overall shapes of its $\sigma_{K}/K$, $K$, and 
$\sigma_{total}$ curves, which are similar to those of the Balmer lines.  The 
values (and $1\sigma$ uncertainies determined from Monte Carlo simulations) of 
the UU Aqr velocity parameters for the Balmer and \ion{He}{1} $\lambda5876$ 
lines are listed in Table~\ref{t-vels} for each of the data sets.
\begin{table}
\dummytable\label{t-vels}
\end{table}

For \ion{He}{1} $\lambda4471$ and \ion{He}{2}, the velocity parameter values 
between $a=1100$ km s$^{-1}$ and $a=1800$ km s$^{-1}$ (where $\sigma_{K}/K$ is 
smallest) oscillate around a roughly flat mean trend, which makes it difficult 
to pick an optimum Gaussian separation.  Instead, we have computed a weighted 
average of the parameter values in this range of $a$ using 
\begin{equation}
\overline{x_{i}} = \frac{\Sigma (x_{i}/\sigma_{i}^{2})}{\Sigma (1/\sigma_{i}^{2})},
\end{equation}
where $\sigma_{i}$ is the uncertainty in parameter $x_{i}$ at a given $a$ as 
determined from a Monte Carlo simulation.  This provides a representative 
velocity solution, rather than relying on any one of the individual solutions 
(which might lie above or below the general trend).  The average parameters for 
these two lines are also listed in Table~\ref{t-vels}.  The Balmer, \ion{He}{1} 
$\lambda4471$ and $\lambda5876$, and \ion{He}{2} $\lambda4686$ radial velocity 
curves obtained from the DIS spectra are shown in Figure~\ref{f-rvel}, along 
with the best fit sine functions.  
For the \ion{He}{1} $\lambda4471$ and \ion{He}{2} lines, we have plotted the 
radial velocities for $a=1700$ km s$^{-1}$ with the sine function produced by 
the average velocity parameters.  The RPCS radial velocity curves for H$\beta$ 
and H$\gamma$ do not differ significantly from the DIS curves, so are not 
plotted.  The He lines in the RPCS data sets were generally too weak to yield 
reliable wing velocity curves. 

All of the lines have positive phase offsets ($\phi_{0}$) corresponding to a 
delay of the red-to-blue velocity crossing of the emission lines relative to 
photometric phase 0.0.  The phase offsets range from $\approx0.1$--$0.2$ in the 
Balmer and \ion{He}{1} lines, to $\approx0.3$ in the \ion{He}{2} line (again, 
however, we cannot rule out the possibility of a skewed result owing to 
blending of the \ion{He}{2} line).  This implies that the main source of 
emission is off-center in the disk.

If we consider the systemic velocities ($\gamma$) as a function of time, then 
the H$\beta$ and H$\gamma$ lines display similar behavior.  In both lines, the 
gamma velocity decreases from data set \#1 to \#2.  In H$\beta$ this decrease 
continues through data set \#3, then $\gamma$ has its maximum value in data set 
\#4; in H$\gamma$ the maximum value is reached in data set \#3.
The value of $\gamma$ then decreases following the maximum for each line, until 
a minimum value ($\gamma<0$ for H$\beta$ and several of the other lines) is 
reached in data set \#7.  
This pattern does not appear to correlate with the smaller 
($\lesssim\pm2\sigma$), somewhat more random fluctuations of the other velocity 
parameters ($K$ and $\phi_{0}$).  This suggests that the systemic velocity is 
influenced by a separate emission region than the velocity semi-amplitude and 
phase offset.  For example, the presence of an emission component from a 
(non-rotating or only slightly rotating) disk wind could systematically skew 
the zero level of the radial velocities without overtly disrupting the 
rotational behavior of underlying emission from the disk.  Variability of the 
flux of emitting material in this wind might then account for the changing 
$\gamma$ values in UU Aqr.

The semi-amplitude of the H$\beta$ radial velocity ranges from $K=77$--$117$ km 
s$^{-1}$, with the best-determined value ($\sigma_{K}/K=0.05$) from the DIS 
data, $K=115\pm6$ km s$^{-1}$.  The variability of the $K$ value may be related 
to small changes in the location and/or structure of the off-center emitting 
region that dominates the Balmer lines.  For the H$\alpha$ line, the DIS 
spectra give $151\pm8$ km s$^{-1}$.  This is slightly smaller than the 
H$\alpha$ wing velocity $K\sim160$ km s$^{-1}$ measured by 
\markcite{Haefner89}Haefner (1989), but larger than that found by 
\markcite{DS91}DS91, $K=121\pm7$ km s$^{-1}$.  
In general, the Balmer lines and (especially) the \ion{He}{2} line have 
velocity semi-amplitudes larger than the nominal WD orbital velocity predicted 
by the photometric model of \markcite{BSC94}BSC94, $K_{wd}=84\pm26$ km 
s$^{-1}$, although many of the H$\beta$ and H$\gamma$ (but not H$\alpha$ or 
\ion{He}{2}) velocities are within $\approx1\sigma$ of the 
\markcite{BSC94}BSC94 value.  The \ion{He}{2} line is often thought to 
originate in the hotter region close to the WD and, hence, to be a good 
indicator of the WD orbital velocity.  However, both its large difference from 
the independently determined value of  \markcite{BSC94}BSC94 and its large 
phase offset argue against adopting our value for $K_{HeII}$ as $K_{wd}$.  
Thus, we will refrain from computing additional system parameters and will 
instead accept those determined for UU Aqr by  \markcite{BSC94}BSC94 from their 
multicolor eclipse mapping investigation (i.e., $q=0.30\pm0.07$, 
M$_{wd}=0.67\pm0.14$M$_{\odot}$, M$_{2}=0.20\pm0.07$M$_{\odot}$, and 
$i=78\pm2^{\circ}$).  We note for the sake of completeness that the effect on 
the derived system parameters of adopting a larger WD orbital velocity would be 
to increase $q$ and decrease $i$ (e.g., \markcite{Garn90}Garnavich et al.\ 
1990).

\subsection{Equivalent Widths}
\label{s-EW}

We measured equivalent widths (EWs) for all 65 of the DIS spectra and for all 
of the spectra from the binned RPCS data sets.  The EW curves for the DIS 
spectra are shown in Figure~\ref{f-EW}; the curves for the RPCS spectra do not 
differ in any significant way from the DIS curves, so are not shown.  The mean 
EWs and $1\sigma$ uncertainties outside of eclipse $(0.1 \le \phi \le 0.9)$ for 
the H$\beta$, H$\gamma$, \ion{He}{1} $\lambda4471$, and \ion{He}{2} 
$\lambda4686$ emission lines of UU Aqr are listed in Table~\ref{t-EW} for each 
of the binned data sets.  
\begin{table}
\dummytable\label{t-EW}
\end{table}
Note that the \ion{He}{2} line is blended with the adjacent 
\ion{C}{3}/\ion{N}{3} emission complex and the EW measures the flux from both.  
The mean EWs of all of these lines were essentially constant from 1995 August 
to 1995 October.  We note, however, that data set \#7 (which has the smallest 
emission line systemic velocities -- see Table~\ref{t-vels}) has the smallest 
mean Balmer and \ion{He}{1} EWs and the largest mean \ion{He}{2} EW.  This 
might be explained if the Balmer and \ion{He}{1} emission contains a variable 
wind emission component that produces a positive systemic velocity when present 
(as discussed in \S\ref{s-rvel}), but is absent or reduced in amplitude in data 
set \#7.

Outside of eclipse, the EWs go through roughly sinusoidal modulations on the 
orbital period, with minima at $\phi\approx$0.7--0.8.  This modulation is most 
apparent in the H$\alpha$ line, where it has a full amplitude of 
$\approx15$\AA\ (see top right panel of Fig.~\ref{f-EW}).  This modulation is 
not seen in broadband (i.e., continuum) light curves of UU Aqr (see 
Fig.~\ref{f-lc} and text below), indicating that it is due to variability in 
the emission lines.  In the novalike CVs BH Lyn (\markcite{Hoard97}Hoard \& 
Szkody 1997) and PG 0859+415 (\markcite{Hoard96a}Hoard \& Szkody 1996a), 
similar modulation of the EWs was interpreted as due to absorption in a 
vertically extended ``bulge'' on the edge of the disk.  The presence of such 
bulges in the disk edge is predicted at phases of 0.2, 0.5, and/or 0.8 by 
numerical simulations of accretion stream-disk interaction (e.g., 
\markcite{Hirose91}Hirose, Osaki, \& Mineshige 1991; 
\markcite{Meglicki93}Meglicki, Wickramasinghe, \& Bicknell 1993).  Absorption 
by a bulge at $\phi\approx0.8$ could explain the UU Aqr EW modulation.

The most obvious feature of the EW curves is the presence of a large peak 
around $\phi\approx0.0$ in all of them.  The height of this peak varies from 
curve to curve; among the Balmer series in particular, its amplitude is largest 
for H$\alpha$ ($\approx$55\AA) and smallest for H$\gamma$ ($\approx$20\AA).  
The eclipse peaks reach maximum amplitude at $\phi=0.0$ but are asymmetrically 
shaped, with a steep rise before $\phi=0.0$ ($\Delta\phi=0.03$--$0.05$) and a 
longer decline ($\Delta\phi=0.05$--$0.10$).  \markcite{DS91}DS91 reported no 
change in the emission line EWs during eclipse, but their spectra were obtained 
in 1988 October, when \markcite{BSC94}BSC94 determined that UU Aqr was in a low 
accretion state.  Yet, the narrowband (50\AA) H$\alpha$ light curve of UU Aqr 
obtained by BSH96 during another low state (in 1992) is uneclipsed relative to 
light curves in both the broadband $R$ and adjacent narrowband $\lambda6700$ 
filters.  This would produce a peak in the H$\alpha$ EW as seen in our high 
state data.  The lack of eclipse in the emission lines implies that there is an 
extended source of emission that remains partially visible during eclipse; 
however, the presence of such a region does not appear to be strongly tied to 
the brightness state in UU Aqr.

We acquired CCD photometry of UU Aqr at SAAO on the nights of 1995 August 
01--02 and 06--07 UT, concurrent with binned RPCS data sets \#1 and \#2 (see 
Table~\ref{t-log}).  A total of 570 measurements were obtained, with typical 
time resolution of $\approx69$ s, and calibrated to Str\"{o}mgren b magnitudes 
via comparison to observations of E-region standards (e.g., 
\markcite{Cousins87}Cousins 1987) obtained under photometric conditions.  The 
eclipse profile in the resultant light curve of UU Aqr (shown in 
Fig.~\ref{f-lc}) is most similar in shape to that of the high state light 
curves of  \markcite{BSC94}BSC94 (compare to their Fig.~5).  The small 
differences in the profile from orbit to orbit immediately before and after 
eclipse are consistent with those seen by \markcite{BSC94}BSC94.  The asymmetry 
of the EW peaks is reflected in the shape of the high state eclipse light 
curve, which shows a steep, smooth ingress to the eclipse minimum over 
$\Delta\phi\approx0.04$, followed by a longer egress, $\Delta\phi\approx0.1$, 
back to the quiescent level.  The eclipse egress has a sharp shoulder 
$\Delta\phi\approx0.04$ after the minimum.   \markcite{BSC94}BSC94 interpreted 
the asymmetry of the eclipse as an effect of the bright spot, which must have a 
large azimuthal extent on the edge of the disk.  The projected surface area of 
the spot along the line-of-sight to the system decreases as the orbital phase 
approaches $\phi=0.0$, leading to a rapid spot eclipse ingress.  After the 
eclipse minimum, the projected spot area along the line-of-sight increases as 
$\phi$ increases, leading to a longer egress from the eclipse of the spot (also 
see \markcite{Rutten92}Rutten et al.\ 1992).  

In addition, the light curve of UU Aqr during 1995 from the Indiana Automated 
CCD Photometric Telescope (``Roboscope;'' e.g., \markcite{Honey92}Honeycutt \& 
Turner 1992, \markcite{Honey94}Honeycutt et al.\ 1994) shows that the system 
was bright ($V=13.5\pm0.1$) during our spectroscopic observations 
(\markcite{Robertson97}Robertson 1997).  As noted in \S\ref{s-intro}, the 
difference in brightness between high and low states of UU Aqr is small, but 
the eclipse profile shapes in the two states show significant differences; 
hence, the latter is a more precise diagnostic of the accretion state in UU Aqr 
than the former.  Based on the comparison of the system brightness, EW 
behavior, and eclipse profile to previously published observations, we conclude 
that UU Aqr was in a high state during our observations.

\subsection{Doppler Tomography}
\label{s-DT}

We used the Fourier-filtered back-projection algorithm described by 
\markcite{Horne91}Horne (1991) and others (e.g., \markcite{Marsh88}Marsh \& 
Horne 1988; \markcite{Kait94}Kaitchuck et al.\ 1994) to produce Doppler 
tomograms of the velocity distribution of emitting material in UU Aqr.  
Figures~\ref{f-HbDT}a--f show the H$\beta$ tomograms for the six binned RPCS 
data sets; Figure~\ref{f-HbDT}g is the H$\beta$ tomogram constructed from the 
DIS data (binned to $\Delta\phi=0.04$); Figure~\ref{f-HbDT}h is the H$\beta$ 
tomogram for the combined set of RPCS spectra.  The tomogram for 1995 Sep 
19--20 UT (Fig.~\ref{f-HbDT}d) is of poorer quality than the others because the 
binned data set used to construct it (\#4 -- see Table~\ref{t-log}) contains 
substantially fewer total spectra than the other RPCS data sets.  As might be 
expected following the similar behavior of the average spectra (see 
\S\ref{s-avsp}) and emission line equivalent widths (\S\ref{s-EW}), the 
H$\beta$ tomogram does not change significantly from 1995 August to 1995 
October.  This implies that the changes in the radial velocity parameters of 
the emission lines must be due to small changes in the structure of a discrete 
emitting region rather than the whole disk.

There are two main features in the tomograms: (1) a roughly circular region of 
diffuse emission, centered around the velocity origin and extending to a radius 
of $\approx$600--800 km s$^{-1}$; and (2) a localized region of strong emission 
centered at $V_{x} \approx -300$ to $-400$ km s$^{-1}$ and $V_{y} \approx 0$ to 
$+50$ km s$^{-1}$.  The latter region is suggestive of the bright spot at the 
impact site of the accretion stream with the edge of the disk, but is somewhat 
inconsistent with the expected velocity position.  Such emission would normally 
lie further inside the $(-V_{x}, +V_{y})$ quadrant of the tomogram (see, for 
example, Fig.\ 11 of \markcite{Kait94}Kaitchuck et al. 1994).  In 
Figures~\ref{f-HbDT}g and \ref{f-HbDT}h, we have plotted the positions of the 
secondary star Roche lobe and the accretion stream trajectory, using the system 
parameters determined by \markcite{BSC94}BSC94 (see \S\ref{s-intro}).  The 
strong emission region does not coincide with the stream trajectory as would be 
expected for normal bright spot emission.  By arbitrarily reducing the WD 
orbital velocity from the value of $\approx84$ km s$^{-1}$ calculated by 
\markcite{BSC94}BSC94 to $\approx40$ km s$^{-1}$, we can force the accretion 
stream to pass roughly through the center of the emission region in the 
tomograms.  However, this requires the masses of the stellar components to be 
unreasonably small ($M_{WD}\approx0.07M_{\odot}$, $M_{2}\approx0.02M_{\odot}$) 
and, consequently, does not provide a useful explanation for the discrepancy 
between stream and emission locations.

The strong emission region also appears to be elongated towards the $(+V_{x}, 
+V_{y})$ quadrant; several of the tomograms (notably a, b, g, and h) show a 
``tail'' of emission which extends from this region to the $+V_{y}$ axis, 
intersecting at $\approx +200$ km s$^{-1}$.  This velocity is comparable to an 
estimate of the orbital velocity of the $L_{1}$ point around the center-of-mass 
assuming the mass ratio calculated by \markcite{BSC94}BSC94.  Indeed, the 
velocity at the end of the emission tail on the $+V_{y}$-axis coincides with 
that of the tip of the secondary star Roche lobe (i.e., the $L_{1}$ point) as 
plotted in Figures~\ref{f-HbDT}g and \ref{f-HbDT}h; the tail itself follows the 
expected trajectory of the accretion stream.
 
None of the H$\beta$ tomograms contains a distinct ring of emission (the 
signature of an accretion disk; e.g., see Fig.\ 24 of 
\markcite{Kait94}Kaitchuck et al. 1994), although there is some indication of 
such a structure in Figures~\ref{f-HbDT}c and \ref{f-HbDT}f.  
Figure~\ref{f-HbDT}g contains the most obvious hint of an emission ring.  The 
lack of a prominent disk signature in the tomograms is not surprising given the 
single- rather than double-peaked emission lines in the spectrum of UU Aqr; 
this may be consistent with only a small disk being present, or only weak 
emission from a larger disk, or emission from a non-disk source (e.g., a wind) 
that masks the disk emission.  
 
The H$\gamma$ and, for the DIS spectra, H$\alpha$ tomograms are very similar to 
the H$\beta$ tomograms (the wavelength coverage of the RPCS spectra does not 
extend to H$\alpha$).  Representative examples of the tomograms of these lines, 
constructed from the DIS data, are shown in Figure~\ref{f-DISdt}a--b.  The 
trailed spectrum used to construct the H$\gamma$ tomogram is shown in 
Figure~\ref{f-trails} (the trailed spectra for H$\alpha$ and H$\beta$ are shown 
in Fig.~\ref{f-linesim}).  There is a somewhat more distinct disk-ring of 
emission in the H$\gamma$ tomogram than in the other Balmer tomograms; the disk 
may be more optically thin to the more energetic Balmer radiation (i.e., at 
hotter temperatures).  The H$\gamma$ ring emission is strongest along an arc 
starting at the strong emission region in the $(-V_{x}, +V_{y})$ quadrant and 
trailing counter-clockwise in the tomogram into the $(+V_{x}, -V_{y})$ 
quadrant, which suggests that the disk edge is nonuniform (in temperature, or 
density, or local turbelent velocity, etc.).

The spectra binned at a given phase from the DIS data set have somewhat higher 
signal-to-noise than the corresponding binned RPCS spectra.  Thus, we also 
attempted to construct Doppler tomograms for the weaker \ion{He}{1} 
$\lambda4471$ and \ion{He}{2} $\lambda4686$ lines in the DIS spectra.  These 
are shown in Figure~\ref{f-DISdt}c--d, and the trailed spectra for these lines 
are shown in Figure~\ref{f-trails}.  The high velocity regions of the 
\ion{He}{1} tomogram are severely contaminated by noise in the continuum 
adjacent to the line.  However, there are still two features of note in the 
tomogram: (1) a region of enhanced emission on the $-V_{x}$ axis, consistent 
with the location of the one seen in the Balmer tomograms; and (2) a 
centralized lack of emission and slightly enhanced arc of emission spanning the 
$-V_{y}$ quadrants that are suggestive of the ring signature of disk emission.  
The \ion{He}{2} tomogram is practically indistinguishable from noise, although 
there may be a slight enhancement of the emission level in the $(-V_{x}, 
+V_{y})$ quadrant (corresponding to the location of the strong emission region 
in the other tomograms).  The poor quality of the \ion{He}{2} tomogram is 
probably the result of both the weakness of the \ion{He}{2} line and its severe 
blending with the adjacent \ion{C}{3}/\ion{N}{3} complex.

The only other published tomogram for UU Aqr is that of the H$\beta$ line shown 
in the CV tomography atlas of \markcite{Kait94}Kaitchuck et al.\ (1994), and it 
is significantly different from our H$\beta$ tomograms.  The strong emission 
region in the Kaitchuck et al.\ tomogram is displaced upward to 
$V_{y}\approx+100$ to $+300$ km s$^{-1}$ and there is a much more prominent 
ring of disk emission (centered at $V_{y}\approx-200$ km s$^{-1}$) than in our 
tomograms.  Overall, the Kaitchuck et al.\ tomogram of UU Aqr is much more 
similar to that of a dwarf nova type CV with a prominent, optically thin disk 
and a discrete bright spot at the expected site of the accretion stream impact 
with the outer edge of the disk.  The Kaitchuck et al.\ tomogram was 
constructed using data obtained in 1988 October, during one of the photometric 
low states classified by  \markcite{BSC94}BSC94.  Thus, we may attribute the 
differences in the tomograms to differences in the accretion state of UU Aqr 
when the spectra were obtained: low for Kaitchuck et al., high for our data.  
The eclipse mapping of  \markcite{BSC94}BSC94 shows that the disk in UU Aqr has 
a larger radius in the high state than in the low; however, the increased mass 
transfer in the high state may produce a strong wind emission component that 
masks the disk emission.  The consistency of our H$\beta$ tomogram from 1995 
August to 1995 October implies that the physical conditions in the UU Aqr disk 
were very stable during the $\approx3$ month span of our observations.

\section{Discussion}

\subsection{Line Profile Simulation}
\label{s-linesim}

As a simple model, CV emission line profiles can be simulated as the sum of 
several component profiles with the appropriate radial velocity vs.\ orbital 
phase behavior expected for various emission regions in the system (e.g., 
accretion stream, bright spot, WD, etc.).  In this manner, 
\markcite{Hellier94}Hellier \& Robinson (1994) and \markcite{Hellier96}Hellier 
(1996) have had some success at reproducing the general line profile behavior 
of SW Sex stars (specifically, PX Andromedae and V1315 Aquilae).  We have 
applied a similar technique to simulating the Balmer emission lines in UU Aqr.
Figure~\ref{f-linesim} shows trailed spectrograms and Doppler tomograms for the 
observed H$\alpha$ (DIS data) and H$\beta$ (combined RPCS data) emission lines 
of UU Aqr, and a simulated Balmer line.  The velocity and phase resolution of 
the simulated data was made comparable to that of the observed data.  Our 
simulated line profiles are made up of five components, described here:

{\bf Double-peaked Keplerian disk emission:} This is the only one of the 
components that is {\em not} represented by a Gaussian profile; instead, the 
disk profile is calculated as in \markcite{Rob93}Robinson, Marsh, \& Smak 
(1993), using an emissivity index $\beta=2.0$ (small changes in $\beta$ around 
this value do not have a significant effect on the simulation owing to the 
relatively small amplitude of the disk profile and the qualitative nature of 
the fitting process -- see below).  We used the system parameters and WD 
orbital velocity for UU Aqr derived by \markcite{BSC94}BSC94 to define the 
shape and orbital phase behavior of the disk profile.  We initially assumed 
inner and outer disk radii of 0.015R$_{\odot}$ and 0.4R$_{\odot}$ 
(\markcite{BSC94}BSC94; \markcite{Harrop96}Harrop-Allin \& Warner 1996).  
Subsequent comparison with the observed profiles (see Fig.~\ref{f-simprof}) 
caused us to pick 0.045R$_{\odot}$ and 0.45R$_{\odot}$ as final inner and outer 
disk radii (the latter value is $\approx60$\% of the distance from the WD to 
the $L_{1}$ point).  These disk radii are used throughout the simulation.  The 
amplitude of the disk profile in the simulation is 0.40 above the continuum 
level (which has been normalized to a value of 1.00 in the observed spectra).

{\bf Single-peaked WD/disk wind emission:} This is represented by a Gaussian 
function with FWHM of 1000 km s$^{-1}$ centered at the midpoint of the disk 
profile.  It is assumed to follow the orbital motion of the WD.  The amplitude 
of the wind component is 0.80 above the continuum.  The amplitudes of both the 
disk and wind components were artificially enhanced by a factor of 2.25 during 
phases 0.98--0.05 to emulate the general appearance of the observed Balmer 
emission lines during eclipse.  We have not explicitly calculated an eclipse 
profile for the simulated line (which would require determination of the 
precise 3-D locations of emission and continuum regions in the system).  Hence, 
the treatment of the eclipse phases shown in the simulated trailed spectrogram 
in Figure~\ref{f-linesim} is, at best, only an approximation of the actual 
eclipse behavior.  (We note, however, that even this simple approximation 
yields line profiles that are similar to the observed profiles during 
eclipse -- see Fig.~\ref{f-simprof} -- although it is not clear whether or not 
this is merely coincidental.)  Since the data from the eclipse phases are 
excluded from the creation of a Doppler tomogram, any incorrect treatment of 
the simulated eclipse behavior has no effect on the simulated tomogram.

{\bf Accretion stream emission:} The free-fall velocities along the accretion 
stream trajectory from the L$_{1}$ point were calculated as in 
\markcite{Lubow75}Lubow \& Shu (1975).  Gaussian profiles with FWHM of 25\% of 
the local stream velocity were summed along a specified length of this 
trajectory to simulate emission from the material in the stream.  In addition, 
the Keplerian rotational velocities of the disk material underlying the stream 
trajectory were calculated, to provide the option of simulating disk material 
excited to emission through interaction with a stream that overflows the disk 
from its initial impact site.  In the case of UU Aqr, we have assumed that the 
stream emits only between the L$_{1}$ point and its impact with the edge of the 
disk -- this reproduces the ``tail'' seen in the Balmer tomograms (see 
\S\ref{s-DT}).  The peak amplitude of the stream component is 0.25 above the 
continuum. 

{\bf Bright spot emission:} In the model proposed by 
\markcite{Hellier94}Hellier \& Robinson (1994) and \markcite{Hellier96}Hellier 
(1996) for the SW Sex stars PX And and V1315 Aql, the accretion stream 
continues coherently past its initial impact with the edge of the disk, 
absorbing the underlying disk emission along its trajectory, to a secondary 
impact site in the inner disk (thereby creating a secondary bright spot whose 
velocity behavior is determined by the dynamics of the inner disk).  The 
eclipse maps of UU Aqr in a high state (\markcite{BSH96}BSH96), however, show 
evidence for a bright spot only at the expected phase of the initial stream 
impact with the disk ($\phi\sim0.8$).  Further, we find that the velocity of a 
secondary impact site near the radius expected for re-impact 
(\markcite{Lubow89}Lubow 1989) must be arbitrarily reduced by a very large 
amount ($>50$\%) to match the observed velocity offset of the strong emission 
region in the Balmer tomograms of UU Aqr.  Consequently, we have simulated the 
line profiles using a bright spot (Gaussian FWHM = 50\% of the average of the 
local stream+disk rotation velocity) that originates at the intersection of the 
stream with the disk edge, and expands to lower velocities (this is consistent 
with, for example, emission from impacting stream material that does not deeply 
penetrate the disk and is primarily reflected away from the disk edge in the 
general direction of the secondary star).  This scenario is suggested by recent 
numerical simulations of the stream-disk interaction 
(\markcite{Armitage96}Armitage \& Livio 1996, \markcite{Armitage97}1997), in 
which the impact of the stream with the disk, in the presence of inefficient 
cooling (appropriate for high $\dot{M}$ systems such as novalike CVs), does 
{\em not} produce coherent stream flow over the disk, but instead causes an 
``explosion'' of material away from the impact point in all directions (except 
those blocked by the disk).  Thus, the stream component in the line simulation 
does not contribute beyond the outer edge of the disk.  The peak amplitude of 
the bright spot profile is 1.20 above the continuum.

{\bf Non-axisymmetric absorption:} \markcite{Hellier97}Hellier (1997) has 
recently proposed accretion stream overflow onto an axisymmetrically flared 
disk to explain the transient absorption features in the emission lines of the 
SW Sex stars.  Unfortunately, this mechanism appears only to be able to 
reproduce the absorption seen around $\phi\approx0.5$ in some of the SW Sex 
stars.  In order to account for absorption at both 0.5 and other phases, there 
must be a non-axisymmetric absorbing structure located at different positions 
in the disk in different systems, as suggested by \markcite{Hoard96a}Hoard \& 
Szkody (1996a, \markcite{Hoard97}1997).  In order to match the observed 
behavior of the Balmer emission lines in UU Aqr,  we utilized non-axisymmetric 
absorption represented by an inverted (i.e., negative amplitude) Gaussian 
function.  The absorbing structure can be visualized as a vertically- and 
azimuthally-extended bulge or wall along the disk edge, which starts (with zero 
absorption) at $\phi=0.95$, ramps up linearly to peak absorption (amplitude of 
$-0.63$ relative to the continuum) between phases 0.7 and 0.8, then ramps back 
down to zero absorption at $\phi=0.3$.  The formation of such a structure could 
result from stream material ``exploding'' around the impact site (near 
$\phi\approx0.8$) to form the highest section of the wall, with additional 
material carried along the disk edge by the rotation of the disk to form the 
long, declining tail of the wall.  This is consistent with the large azimuthal 
extent of the bright spot expected from the eclipse profiles of UU Aqr (see 
\S\ref{s-EW}).  The FWHM of the absorption is 300 km s$^{-1}$; this value was 
initially set at the rotational velocity of the outer edge of the disk 
($\approx550$ km s$^{-1}$) and was refined by comparing the simulated 
absorption width with that seen between phases 0.5 and 1.0 in the observed 
spectra.  In the simulation shown in Figure~\ref{f-linesim}, the velocity 
behavior of the absorption was set to follow that of the bright spot, in 
essence producing a self-absorption component of the bright spot (as suggested 
by \markcite{Dick97}Dickinson et al. 1997 to explain the line profile behavior 
of the probable SW Sex star V795 Herculis).  We also experimented with somewhat 
more physical forms of velocity behavior for the absorption, such as following 
either the rotational velocity of the edge of the disk or the continuation of 
the stream trajectory over the disk, and found these to produce qualitatively 
similar results.  Considering the simplicity of our simulation, we elected to 
present here the case involving the fewest unknown parameters (i.e., 
self-absorption).

In general, this technique of line profile simulation can produce 
reasonable-looking results.  Yet, the large number of input parameters and use 
of simplifying approximations beg the question: how {\em realistic} are these 
results?  We have calculated the $\chi^{2}$ values of the simulated profiles 
using different input parameters compared to the observed profiles, in a number 
of phase bins; the comparison for the simulation presented here is shown in 
Figure~\ref{f-simprof}.  This provided a means of comparing the relative 
``goodness'' of simulations for a particular data set.  However, the parameters 
were adjusted based only on a visual inspection of the simulated vs.\ real 
profiles; for example, the blue shoulder visible in the $\phi=0.15$ profile in 
Figure~\ref{f-simprof} provided a measure of the correct amplitudes for the 
disk and wind components, while the height of the central peak provided an 
estimate of the correct amplitude for the bright spot emission.  This manual 
fitting can produce somewhat erratic results.  Further, like Doppler 
tomography, the line profile simulation process essentially collapses the 
velocity field of the emission into two dimensions, which can lead to some 
ambiguity if significant sources of emission have large velocity components in 
the third dimension (i.e., vertical relative to the plane of the disk).

\subsection{Half-Orbit Tomography}
\label{s-halfDT}

The presence of a non-axisymmetric absorbing structure in the disks of SW Sex 
stars appears to be quite successful at explaining the transient absorption 
features in the emission lines of these CVs.  Unfortunately, it poses a 
potential problem for the technique of Doppler tomography, which is widely used 
in the analysis of both the SW Sex stars and CVs in general.  A basic tenet of 
the tomography process is that the emission region(s) being mapped should be 
equally visible at all orbital phases of the system; this is why eclipse phase 
data are neither used in the tomography process nor accurately reproduced in 
forward-projections of the velocity maps.  In order to circumvent the violation 
of this tenet in UU Aqr, we have constructed two additional sets of Balmer line 
tomograms.  In the first set (shown in the top panels of Fig.~\ref{f-halfDT}), 
we used only the spectra from orbital phases 0.0 to 0.5 -- the stronger, 
``unabsorbed'' spectra (see top panels of Fig.~\ref{f-linesim}).  The second 
set uses the remaining, ``absorbed'' spectra, from $\phi=0.5$--$1.0$ (bottom 
panels of Fig.~\ref{f-halfDT}).  Thus, each of these half-orbit time-resolved 
tomograms should show the emission from only one side of the absorbing 
structure described in \S\ref{s-linesim}.  

In the $\phi=0.0$--$0.5$ tomograms, we are effectively looking across most of 
the width of the disk to the absorbing structure on the far edge.  The strong 
emission region lies on the $-V_{x}$ axis in the same location as in the 
full-orbit tomograms.  This emission originates at the explosive stream impact 
with the disk edge, and is viewed on the inner disk side of the absorbing 
structure.  Little or no emission is visible along the accretion stream 
trajectory itself; the stream is likely obscured at most of these phases by the 
absorbing bulge, accretion disk, and/or secondary star.  In the 
$\phi=0.5$--$1.0$ tomograms, we are looking only at the stream and the small 
amount of disk material at radii outside the absorbing structure.  There is 
prominent emission along the accretion stream trajectory; however, there is 
little or no emission on the $-V_{x}$ axis.  Presumably, most of the stream 
impact emission is obscured by the absorbing structure at these phases.  

It was noted in \S\ref{s-DT} that the H$\gamma$ tomogram showed the most 
prominent suggestion of the ring indicative of an accretion disk.  The 
half-orbit tomograms of this line show that the strongest part of this ring 
emission is visible only at $\phi=0.5$--$1.0$, in the form of enhanced emission 
that trails out along the disk-ring from the bright spot at the stream impact 
site.  This corresponds to material carried along the outer edge of the disk by 
the disk rotation (e.g., \markcite{Armitage97}Armitage \& Livio 1997); it is 
hidden behind the absorbing structure and/or secondary star for 
$\phi\approx0.0$--$0.5$, so is not seen in the $\phi=0.0$--$0.5$ tomograms.  A 
similar trail of emission is seen in tomograms of the archetype SW Sex 
(\markcite{Dhillon97}Dhillon, Marsh, \& Jones 1997) and the non-SW Sex system 
WZ Sagittae (\markcite{Spruit97}Spruit \& Rutten 1997);  Spruit \& Rutten 
suggest that this emission is due to recombination in material that is cooling 
down from the shock of the stream-disk impact as it is swept ``downstream'' by 
the disk rotation.

\subsection{A System Model}
\label{s-model}

Figure~\ref{f-diag} is a schematic diagram of a CV showing the features of the 
model for UU Aqr suggested by our line profile simulations and half-orbit 
tomography.  The high $\dot{M}$ accretion stream strikes the edge of the disk 
and forms a roughly spherical ``explosion'' of emitting material (region A in 
Fig.~\ref{f-diag}).   This emitting material is flung primarily in directions 
of least resistance, away (and above) the disk, but some also disperses across 
the inner disk, producing diffuse emission in the tomograms (region B).  An 
optically thick absorbing wall is built up by large amounts of the impacting 
stream material (region D), either carried along the disk edge or roughly 
following the continuation of the stream trajectory (although there is no 
coherent overflow of the stream).  Optically thin emitting gas heated by the 
shock of the stream impact is also carried along the very outer edge of the 
disk by the disk rotation, outside the absorbing structure (relative to the 
WD), forming the trailing arc of emission seen most prominently in the 
H$\gamma$ tomograms (region C).  (This situation is likely to be further 
complicated by the probable presence of a wind from the disk that is mentioned, 
but not explicitly addressed, in this work.)

\subsection{UU Aqr as an SW Sex Star}
\label{s-disc}

The definition of the SW Sex stars as a distinct class of CV is somewhat 
ambiguous and considerably more phenomenological than that of other CV classes 
(e.g., U Geminorum stars, polars, etc.)  The SW Sex stars tend to be high 
inclination systems, but this is almost certainly a selection effect.  They 
have similar orbital periods, $P_{orb}\approx3$--$4$ h, which suggests a 
possible evolutionary link to the onset of SW Sex behavior.  In general, they 
display single-peaked emission lines, orbital-phase-dependent absorption in the 
Balmer lines, and phase offsets in the emission line radial velocity curves.  
UU Aqr has all of these general qualities.  Early work on the SW Sex stars 
(e.g., \markcite{Szkody90}Szkody \& Pich\'{e} 1990; 
\markcite{Thorst91}Thorstensen et al. 1991) specified that the absorption 
occurred only at $\phi\approx0.5$, whereas it occurs at $\phi\approx0.8$ in UU 
Aqr (and in PG 0859+415).  Yet, this variation in behavior can be accommodated 
if the SW Sex ``class'' contains systems comprising a continuum of 
observational characteristics determined by the relative values of fundamental 
parameters (such as the mass transfer rate), rather than many examples of a 
fixed system morphology.

We note that strong \ion{He}{2} $\lambda4686$ emission (relative to the Balmer 
emission lines, for example) is also typically mentioned as a distinguishing 
property of the SW Sex stars, but UU Aqr has weak \ion{He}{2}.  The strength of 
\ion{He}{2}, however, does vary among the members of this class 
(\markcite{Hoard96b}Hoard \& Szkody 1996b), from comparable to H$\beta$ (e.g., 
DW UMa; \markcite{Shafter88}Shafter et al.\ 1988) to much weaker than H$\beta$ 
(e.g., WX Arietis; \markcite{Beuer92}Beuermann et al.\ 1992).  Thus, although 
the reason for the weak \ion{He}{2} emission in UU Aqr is not entirely clear, 
just the fact that it is weak is not so troubling.  It is reasonable to suggest 
that the strength of \ion{He}{2} is related to inclination; for example, WX Ari 
($i\lesssim75^{\circ}$; \markcite{Beuer92}Beuermann et al.\ 1992) and PG 
0859+415 ($i\approx65^{\circ}$; \markcite{Hoard96a}Hoard \& Szkody 1996a), 
which are both moderate inclination systems, have weak \ion{He}{2} emission.  
However, this correlation does not appear to extend to other systems: BH Lyn 
($i=79^{\circ}$; \markcite{Hoard97}Hoard \& Szkody 1997) and SW Sex 
($i=79^{\circ}$; \markcite{Penning84}Penning et al.\ 1984) have inclinations 
similar to that of UU Aqr but have strong \ion{He}{2} emission, while PX And 
($i=73.6^{\circ}$; \markcite{Thorst91}Thorstensen et al.\ 1991) has a smaller 
inclination than UU Aqr, but stronger \ion{He}{2} emission.  The presence of 
optically thick parts of the outer disk that obscure the hot inner regions 
where \ion{He}{2} is expected to form might reduce the emission strength; for 
example, PG 0859+415 and UU Aqr are known or suspected of having thick disks 
and both display weak \ion{He}{2} emission.

The dependency of the amount of stream overflow on cooling efficiency noted by 
\markcite{Armitage97}Armitage \& Livio (1997) provides at least a partial 
explanation for the range of behavior among the SW Sex stars.  At high mass 
transfer rates, cooling is inefficient and the stream impact produces a 
forceful explosion, resulting in no coherent stream overflow and the presence 
of line absorption at the initial impact site ($\phi\approx0.8$).  At low mass 
transfer rates, cooling is efficient and there is substantial, coherent 
overflow of the stream.  This allows the formation of a bright spot in the 
inner disk at the site of the secondary stream impact site, resulting in line 
absorption around $\phi\approx0.5$.  In the SW Sex stars UU Aqr and PG0859+415 
(\markcite{Hoard96a}Hoard \& Szkody 1996a), the line absorption is deepest near 
$\phi=0.8$ and there is evidence for a substantial bright spot on the edge of 
the disk.  In BH Lyn, the absorption occurs around $\phi=0.5$ but there is also 
evidence for the presence of a bright spot on the edge of the disk from time to 
time (\markcite{Hoard97}Hoard \& Szkody 1997).  In DW UMa 
(\markcite{Shafter88}Shafter et al.\ 1988), PX And 
(\markcite{Thorst91}Thorstensen et al.\ 1991), and V1315 Aql 
(\markcite{DMJ91}Dhillon, Marsh, \& Jones 1991), the absorption occurs around 
$\phi=0.5$ and there is little or no evidence for a bright spot on the disk 
edge.  We can understand this range of behavior in the context of the numerical 
simulations of \markcite{Armitage97}Armitage \& Livio (1997) if the mass 
transfer rate is largest for UU Aqr and PG 0859+415, $\dot{M}$ is smaller for 
BH Lyn, and still smaller for DW UMa, PX And, and V1315 Aql.  This produces 
primarily explosive impact of the stream with the disk in the first two 
systems, a mix of explosive impact and stream overflow in BH Lyn, and primarily 
stream overflow in the last three systems.  Mass transfer rates in CVs are 
notoriously difficult to estimate, but we note that if their $\dot{M}$'s do 
compare as described here, then the orbital periods of these six systems (UU 
Aqr = 3.9 h; PG 0859+415 = 3.7 h; BH Lyn = 3.7 h; PX And = 3.5 h; V1315 Aql = 
3.3 h; DW UMa = 3.3 h) roughly follow a trend where mass transfer rate 
decreases with decreasing orbital period.  Some evidence for such a trend has 
been seen among the novalike CVs (\markcite{Dhillon96}Dhillon 1996). 

\acknowledgments

We wish to thank Keith Horne for making his Doppler tomography software 
available to us, and Stefanie Wachter for reading a draft of this paper.  The 
research of PS and DWH was supported by NASA grant NAG-W-3158 and NSF grant 
AST9217911.  MDS was supported by the Particle Physics and Astronomy Research 
Council grant K46019.  The work of RCS was partially supported by PPARC grant 
GR/K45555.

\newpage
\begin{center}{\bf FIGURE CAPTIONS}\end{center}
\figcaption{Average red (bottom) and blue (top) spectrum of UU Aqr, uncorrected 
for orbital motion.  The red spectrum is the DIS data from 1995 Oct 12--13 UT.  
The blue spectra are from (top to bottom) 1995 Aug 1--3 UT, 1995 Aug 6--8 UT, 
1995 Aug 18--20 UT, 1995 Sep 19--20 UT, 1995 Sep 21-23 UT, 1995 Sep 24--26 UT, 
1995 Oct 12--13 UT.  The blue flux scale refers to the bottom spectrum (DIS 
data); the other spectra have been cumulatively offset by $0.5\times10^{-14}$ 
erg s$^{-1}$ cm$^{-2}$ \AA$^{-1}$.\label{f-avsp}}

\figcaption{Diagnostic diagrams from the double Gaussian fitting technique for the 
DIS data. The left panels are for the Balmer lines: H$\alpha$ (dotted line), 
H$\beta$ (solid line), and H$\gamma$ (dashed line).  The right panels are for 
the helium lines: He~{\sc i} $\lambda4471$ (dotted line), He~{\sc ii} 
$\lambda4686$ (solid line), and He~{\sc i} $\lambda5876$ (dashed line).  The 
panels show (from bottom to top) the phase shift of the velocity solution, the 
systemic velocity (km s$^{-1}$), the fractional uncertainty of the velocity 
semi-amplitude, the velocity semi-amplitude (km s$^{-1}$), and the total 
$1\sigma$ uncertainty of the velocity fit (km s$^{-1}$).\label{f-diags}}

\figcaption{Emission line radial velocity curves measured from the wings of the 
DIS spectra using the double Gaussian fitting technique.\label{f-rvel}}

\figcaption{Equivalent width curves of the prominent emission lines in the DIS 
spectra of UU Aqr.  The top right panel shows the same H$\alpha$ data as the 
top left panel plotted on an expanded scale to show the variation of the 
equivalent width outside of eclipse.\label{f-EW}}

\figcaption{Light curve of UU Aqr obtained on 1995 August 01--02 and 06--07 UT.  
The top panel shows the entire curve, phased according to the ephemeris of  
BSC94, while the bottom panel shows a close-up of the region around the 
eclipse, with data obtained during different orbital cycles plotted with 
differently-shaped points.\label{f-lc}}

\figcaption{Doppler tomograms for the H$\beta$ emission line of UU Aqr from 1995 
August to 1995 October.  Panels (a--f) were constructed using the binned RPCS 
data sets from (a) 1995 Aug 1--3 UT, (b) 1995 Aug 6--8 UT, (c) 1995 Aug 18--20 
UT, (d) 1995 Sep 19--20 UT, (e) 1995 Sep 21--23 UT, and 
(f) 1995 Sep 24--26 UT.  Panel (g) was constructed using the binned DIS data 
set.  Panel (h) was constructed using the combined set of RPCS spectra.  The 
secondary star Roche lobe and accretion stream trajectory are plotted in panels 
(g) and (h) using the system parameters determined by BSC94.\label{f-HbDT}}

\figcaption{Doppler tomograms for several emission lines constructed from the DIS 
data: (a) H$\alpha$, (b) H$\gamma$, (c) \ion{He}{1} $\lambda4471$, (d) 
\ion{He}{2} $\lambda4686$ (+ \ion{C}{3}/\ion{N}{3}).\label{f-DISdt}}

\figcaption{Trailed spectra for the He~{\sc ii} + C~{\sc iii}/N~{\sc iii} emission 
complex (left), the He~{\sc i} $\lambda4471$ emission line (middle), and the 
H$\gamma$ emission line (right).  Emission is dark; continuum is light.  The 
spectra are repeated over two orbital cycles.  Note that the velocity scale of 
the He~{\sc ii} panel is larger than that of the other two panels to show the 
adjacent C~{\sc iii}/N~{\sc iii} emission.\label{f-trails}}

\figcaption{Trailed spectrograms and Doppler tomograms of UU Aqr.  The left panels 
show the H$\alpha$ line from the DIS spectra.  The right panels show the 
H$\beta$ line from the total combined RPCS data set.  The middle panels show 
simulated data.  The Roche lobe of the secondary star, the path of the 
accretion stream (lower curve), and the position of Keplerian disk velocities 
under the stream trajectory (upper curve) have been plotted in the Doppler 
tomograms, based on the system parameters determined by BSC94.\label{f-linesim}}

\figcaption{Observed (solid line) and simulated (dotted line) 
orbital-phase-resolved emission line profiles for UU Aqr.  The observed data is 
the H$\alpha$ line from the DIS spectra.\label{f-simprof}}

\figcaption{Half-orbit tomograms for the H$\alpha$ (left panels), H$\beta$ (middle 
panels), and H$\gamma$ (right panels) emission lines of UU Aqr.  The DIS data 
were used for H$\alpha$ and H$\gamma$; the RPCS data for H$\beta$.  The 
tomograms in the top panels were constructed using only the spectra in the 
range $0.0 \le \phi \le 0.5$; those in the lower panels used only the spectra 
in the range $0.5 \le \phi \le 1.0$.  The linear structures visible in the 
tomograms are artifacts caused by the decreased orbital phase coverage of the 
input spectra.\label{f-halfDT}}

\figcaption{A schematic model of UU Aqr.  The Roche-lobe-filling secondary star is 
shown on the right, and the white dwarf (black dot) and accretion disk are on 
the left.  The orbital phases of a number of lines-of-sight through the system 
are labeled around the perimeter of the dashed circle enclosing the diagram.  
Phase 0.0 corresponds to the superior conjunction of the WD.  The system 
rotates counter-clockwise in its orbit, which corresponds to an observer moving 
clockwise around the lines of sight down each fiducial phase.  The lettered 
regions are described in \S4.3 of the text.\label{f-diag}}

\begin{deluxetable}{ccclcccc}
\tablecolumns{8}
\tablewidth{0pt}
\tablecaption{Observation Log}
\tablenum{1}
\tablehead{
\colhead{Binned} &
\colhead{mid-JD of} &
\colhead{ } &
\colhead{ } &
\multicolumn{2}{c}{UT Time} &
\colhead{Orbital} &
\colhead{\# of} \\
\colhead{Data Set} &
\colhead{Binned Data} &
\colhead{Instrument} &
\colhead{UT Date} &
\colhead{Start} &
\colhead{End} &
\colhead{Phases} &
\colhead{Spectra} 
}
\startdata
1 & 2449932 & RPCS & 1995 Aug 01--02 & 21:24 & 02:04 & 0.51--1.70 & 161 \\
  &         & RPCS & 1995 Aug 02--03 & 22:16 & 02:28 & 0.84--1.91 & 146 \\ \\
2 & 2449937 & RPCS & 1995 Aug 06--07 & 21:33 & 02:59 & 0.11--1.50 & 187 \\
  &         & RPCS & 1995 Aug 07--08 & 21:58 & 03:00 & 0.33--1.61 & 172 \\ \\
3 & 2449949 & RPCS & 1995 Aug 18--19 & 22:44 & 02:40 & 0.77--1.78 & 145 \\
  &         & RPCS & 1995 Aug 19--20 & 22:43 & 02:18 & 0.88--1.80 & 122 \\ \\
4 & 2449980 & RPCS & 1995 Sep 19--20 & 23:20 & 01:21 & 0.55--1.06 & \phn69 \\
  &         & RPCS & 1995 Sep 20     & 20:05 & 22:58 & 0.83--1.57 & \phn91 \\ \\5 & 2449982 & RPCS & 1995 Sep 21--22 & 20:16 & 00:37 & 0.99--2.10 & 146 \\
  &         & RPCS & 1995 Sep 22--23 & 23:13 & 00:56 & 0.86--1.29 & \phn58 \\ \\6 & 2449984 & RPCS & 1995 Sep 24--25 & 21:43 & 01:05 & 0.70--1.56 & 111 \\
  &         & RPCS & 1995 Sep 25--26 & 22:40 & 00:58 & 0.06--1.64 & \phn84 \\ \\7 & 2450003 & DIS  & 1995 Oct 12     & 04:45 & 08:18 & 0.41--1.31 & \phn35 \\
  &         & DIS  & 1995 Oct 13     & 05:06 & 08:07 & 0.61--1.38 & \phn30 
\enddata
\end{deluxetable}

\begin{deluxetable}{ccccccccccc}
\tablecolumns{13}
\tablewidth{0pt}
\tablecaption{Radial Velocity Parameters from Emission Line Wings}
\tablenum{2}
\tablehead{
\colhead{$\!$Data} &
\multicolumn{5}{l}{$\!\!$H$\beta$:} &
\multicolumn{5}{l}{$\!\!$H$\gamma$:} \\
\colhead{$\!$Set} &
\colhead{$\!\!\gamma\,$\tablenotemark{a}} &
\colhead{$\!\!\!\!K$} &
\colhead{$\!\!\!\phi_{0}$} &
\colhead{$\!\!\!\sigma_{K}/K$} &
\colhead{$\!\!\!\sigma_{TOTAL}$} &
\colhead{$\!\!\gamma\,$\tablenotemark{a}} &
\colhead{$\!\!\!\!K$} &
\colhead{$\!\!\!\phi_{0}$} &
\colhead{$\!\!\!\sigma_{K}/K$} &
\colhead{$\!\!\!\sigma_{TOTAL}$} \\
\colhead{$\!\!$ } &
\colhead{$\!\!$(km s$^{-1}$)} &
\colhead{$\!\!\!\!$(km s$^{-1}$)} &
\colhead{$\!\!\!$ } &
\colhead{$\!\!\!$ } &
\colhead{$\!\!\!$(km s$^{-1}$)} &
\colhead{$\!\!$(km s$^{-1}$)} &
\colhead{$\!\!\!\!$(km s$^{-1}$)} &
\colhead{$\!\!\!$ } &
\colhead{$\!\!\!$ } &
\colhead{$\!\!\!$(km s$^{-1}$)} 
}
\startdata
$\!\!\!$1 & 36(2) & $\!\!\!\!$117(12) & 0.15(2) & 0.10 & $\!\!\!\!$43 & 27(4) & 
119(8)\phn & 0.15(2) & 0.07 & $\!\!\!\!$36 \\
$\!\!\!$2 & \phn8(1) & $\!\!\!\!$113(18) & 0.14(2) & 0.16 & $\!\!\!\!$59 & 15(3)
& 107(15) & 0.11(2) & 0.14 & $\!\!\!\!$48 \\
$\!\!\!$3 & \phn4(4) & $\!\!\!\!$114(12) & 0.16(2) & 0.11 & $\!\!\!\!$50 & 81(5)
& \phn97(11) & 0.18(2) & 0.11 & $\!\!\!\!$45 \\
$\!\!\!$4 & 86(5) & $\!\!\!\!$\phn95(11) & 0.13(3) & 0.12 & $\!\!\!\!$45 &
\phn65(10) & 117(22) & 0.12(4) & 0.19 & $\!\!\!\!$78 \\
$\!\!\!$5 & \phn8(6)\phn & $\!\!\!\!$\phn77(12) & 0.10(3) & 0.16 & $\!\!\!\!$49 
& 45(4) & 121(19) & 0.15(3) & 0.16 & $\!\!\!\!$67 \\
$\!\!\!$6 & 31(4) & $\!\!\!\!$113(16) & 0.09(2) & 0.14 & $\!\!\!\!$51 & 31(8)
& 104(12) & 0.08(3) & 0.11 & $\!\!\!\!$51 \\
$\!\!\!$7 & $-$19(1)\phs & $\!\!\!\!$115(6)\phn & 0.13(1) & 0.05 & $\!\!\!\!$34 
& 18(2) & 103(8)\phn & 0.10(1) & 0.08 & $\!\!\!\!$43 \\ \\
  & \multicolumn{5}{l}{H$\alpha$:} & \multicolumn{5}{l}{He~{\sc i} 
$\lambda5876$:} \\
$\!\!\!$7 & \phn$-$4(1)\phs   & $\!\!\!\!$151(8)\phn    & 0.21(1) & 0.05 & 
$\!\!\!\!$38 & \phn$-$58(1)\phs  & \phn61(7)\phn & 0.16(2) & 0.12 & $\!\!\!\!$37
\\ \\
  & \multicolumn{5}{l}{He~{\sc i} $\lambda4471\,$: \tablenotemark{c}} & 
\multicolumn{5}{l}{He~{\sc ii} $\lambda4686\,$: \tablenotemark{b,c}} \\
$\!\!\!$7 & 33\phn\phn & $\!\!\!\!$156\phn\phn\phn & 0.13\phn\phn & 0.16 & 
$\!\!\!\!$135\phn & $-$87\phn\phn & 226\phn\phn\phn & 0.27\phn\phn & 0.15 & 
$\!\!\!\!$165\phn 
\enddata
\tablenotetext{a}{Heliocentric corrections applied to systemic velocities.}
\tablenotetext{b}{Blended with C~{\sc iii}/N~{\sc iii} emission.}
\tablenotetext{c}{Weighted average of velocity parameter values in the range 
$a=1100$--$1800$ km s$^{-1}$ -- see text.}
\end{deluxetable}
\begin{deluxetable}{clcccc}
\tablecolumns{6}
\tablewidth{0pt}
\tablecaption{Mean Emission Line Equivalent Widths $(0.1\le\phi\le0.9)$}
\tablenum{3}
\tablehead{
\colhead{Data} &
\colhead{ } &
\multicolumn{4}{c}{Mean Equivalent Width (\AA):} \\
\colhead{Set} &
\colhead{UT Date} &
\colhead{H$\beta$} &
\colhead{H$\gamma$} &
\colhead{He~{\sc i} $\lambda4471$} &
\colhead{He~{\sc ii} $\lambda4686\,$\tablenotemark{a}} 
}
\startdata
1 & 1995 Aug 01--03 & 18.2(1.1) & 12.1(2.0) & 1.4(0.7) & 9.8(1.6) \\
2 & 1995 Aug 06--08 & 18.3(2.3) & 12.7(1.4) & 2.1(0.8) & 9.7(1.4) \\
3 & 1995 Aug 18--20 & 21.1(3.4) & 11.8(2.5) & 2.1(1.4) & 6.7(1.8) \\
4 & 1995 Sep 19--20 & 19.0(3.0) & 13.3(7.5) & 2.0(1.9) & 8.2(5.7) \\
5 & 1995 Sep 21--23 & 21.9(3.6) & 13.9(2.2) & 2.6(1.2) & 4.6(5.5) \\
6 & 1995 Sep 24--26 & 17.6(2.1) & 12.1(3.2) & 2.7(1.2) & 8.5(1.8) \\
7 & 1995 Oct 12--13 & 17.2(2.8) & \phn9.2(2.5)  & 0.0(1.0) & 10.2(1.9)\phn \\
\enddata
\tablenotetext{a}{Blended with C~{\sc iii}/N~{\sc iii} emission.}
\end{deluxetable}

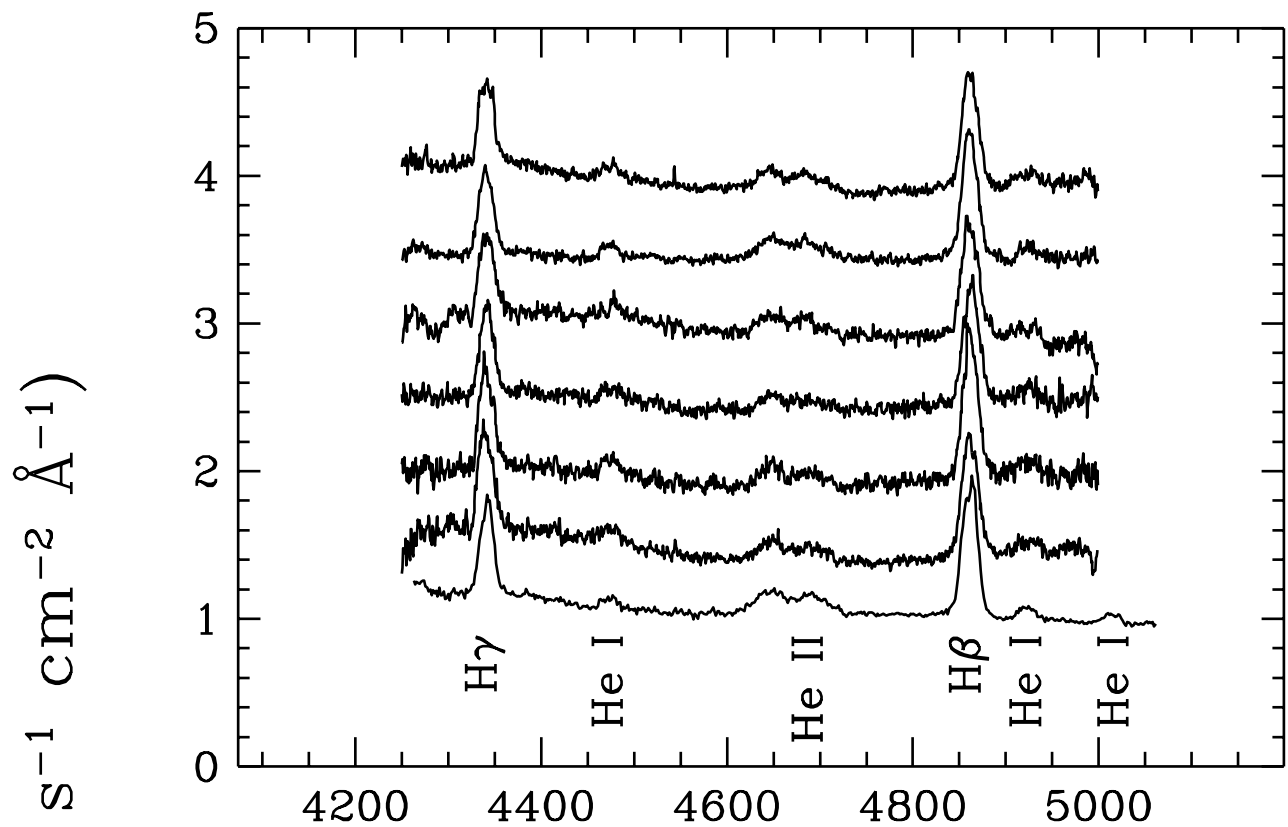
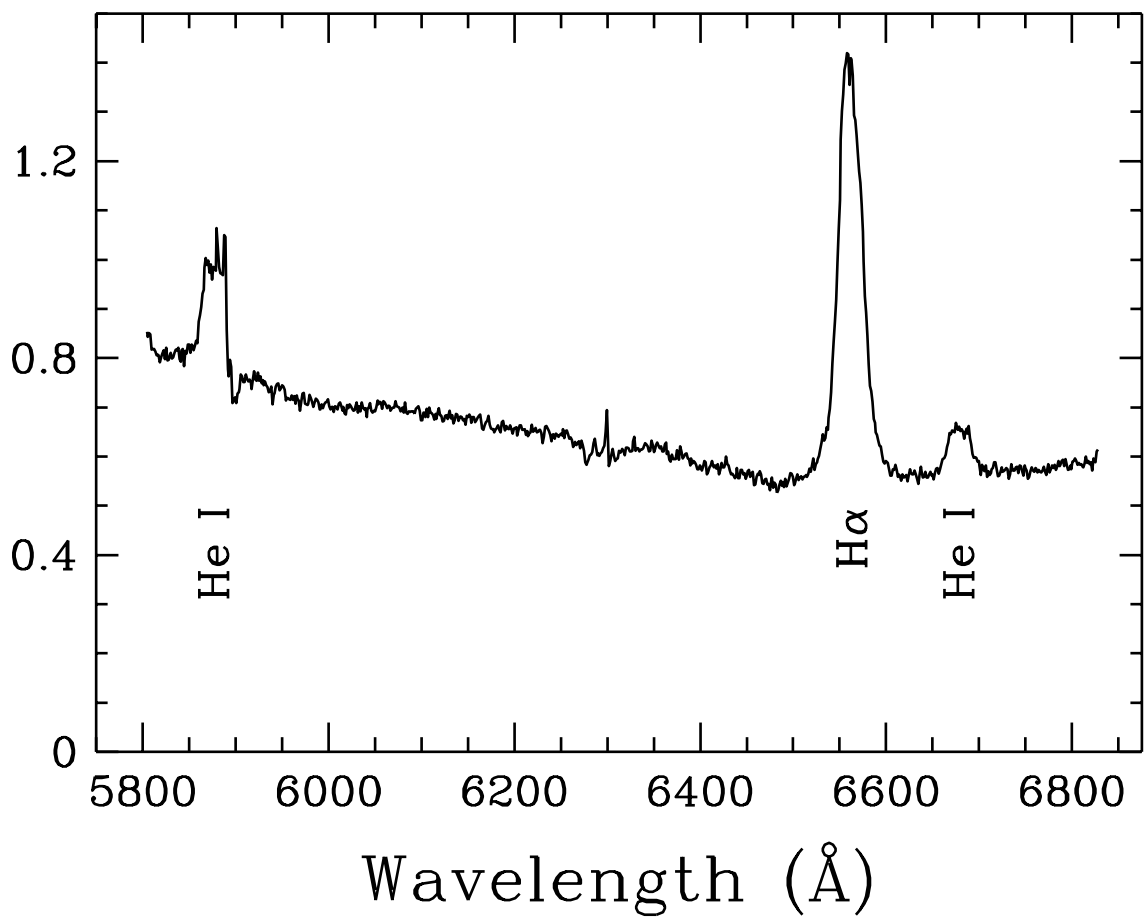
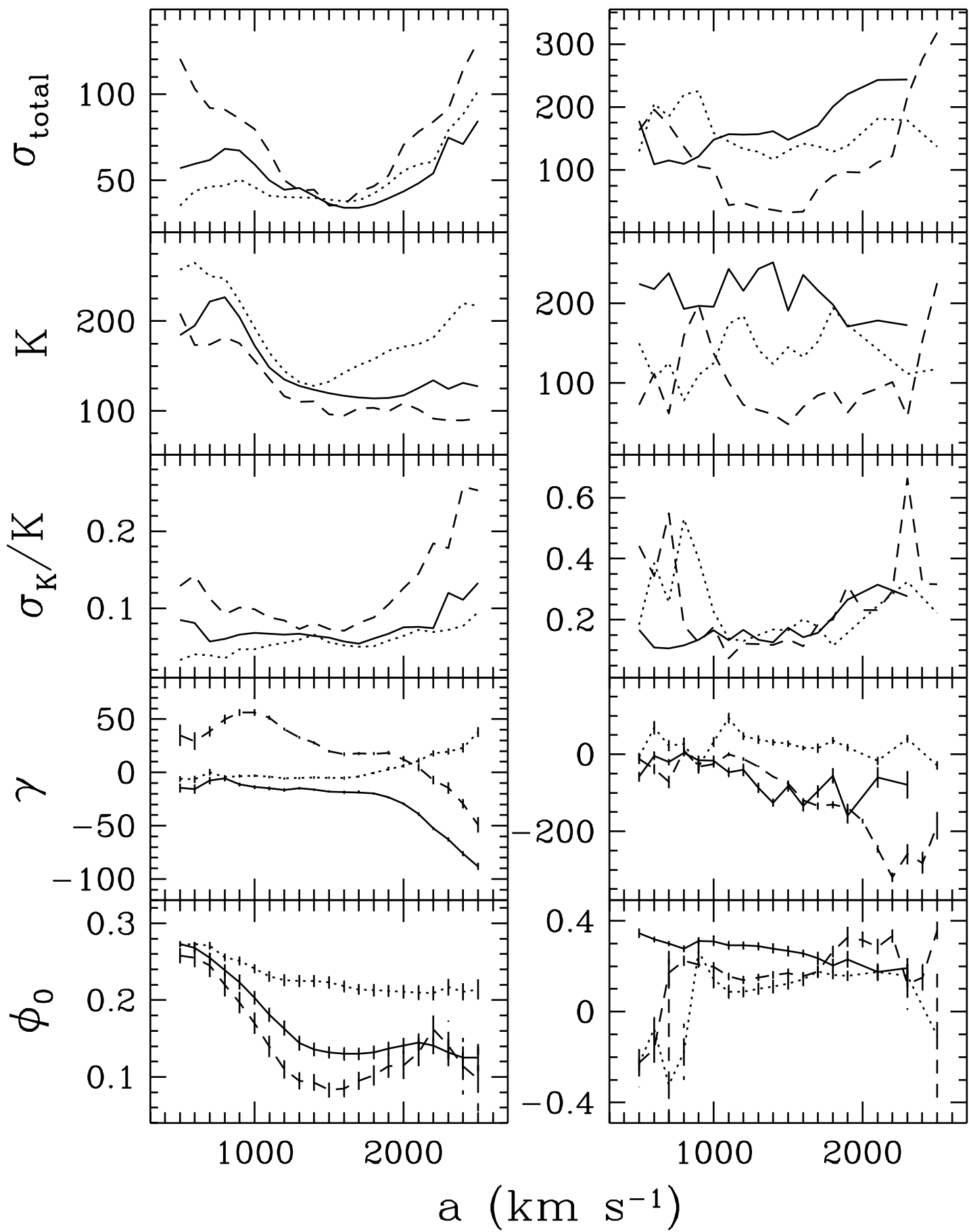
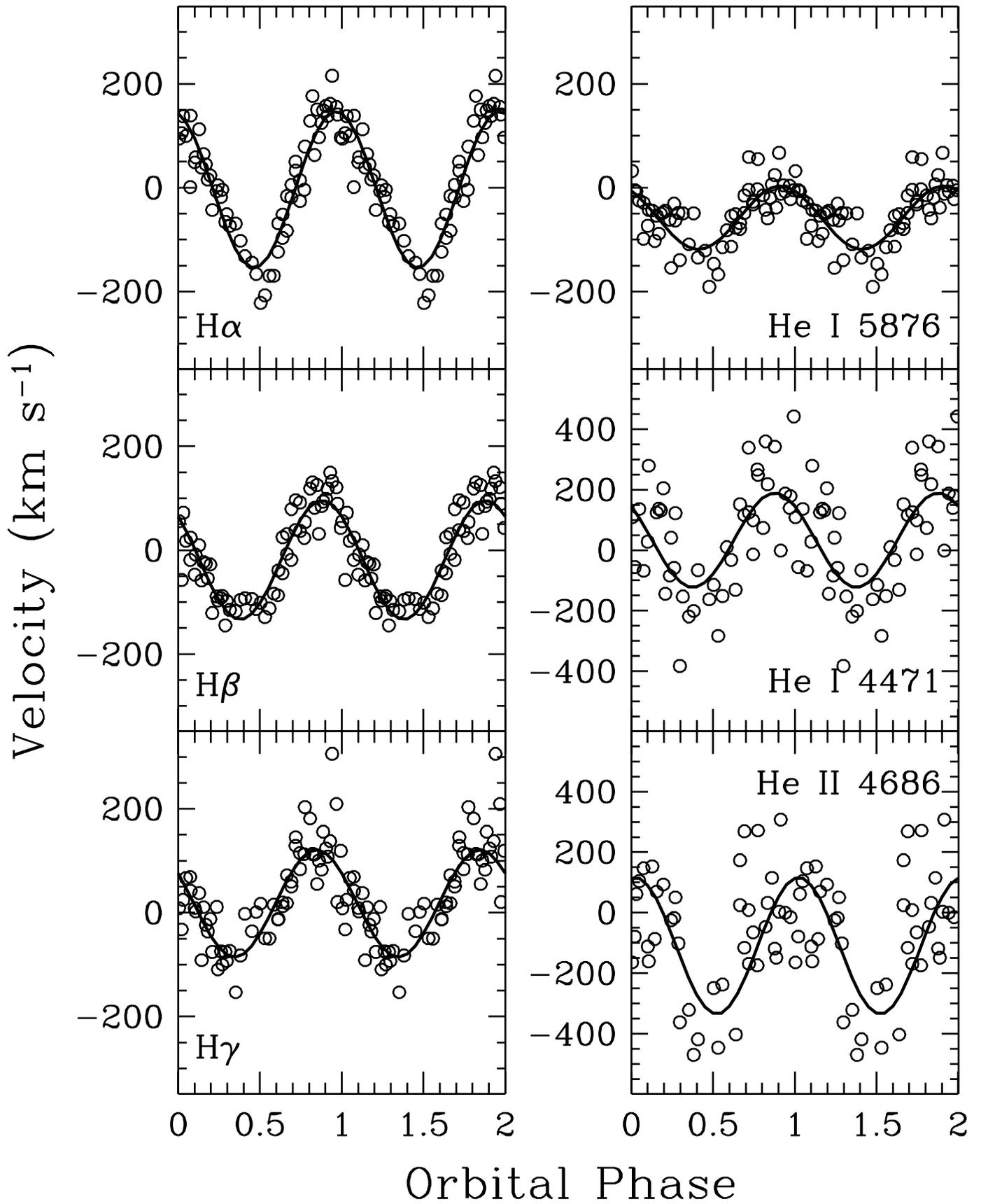
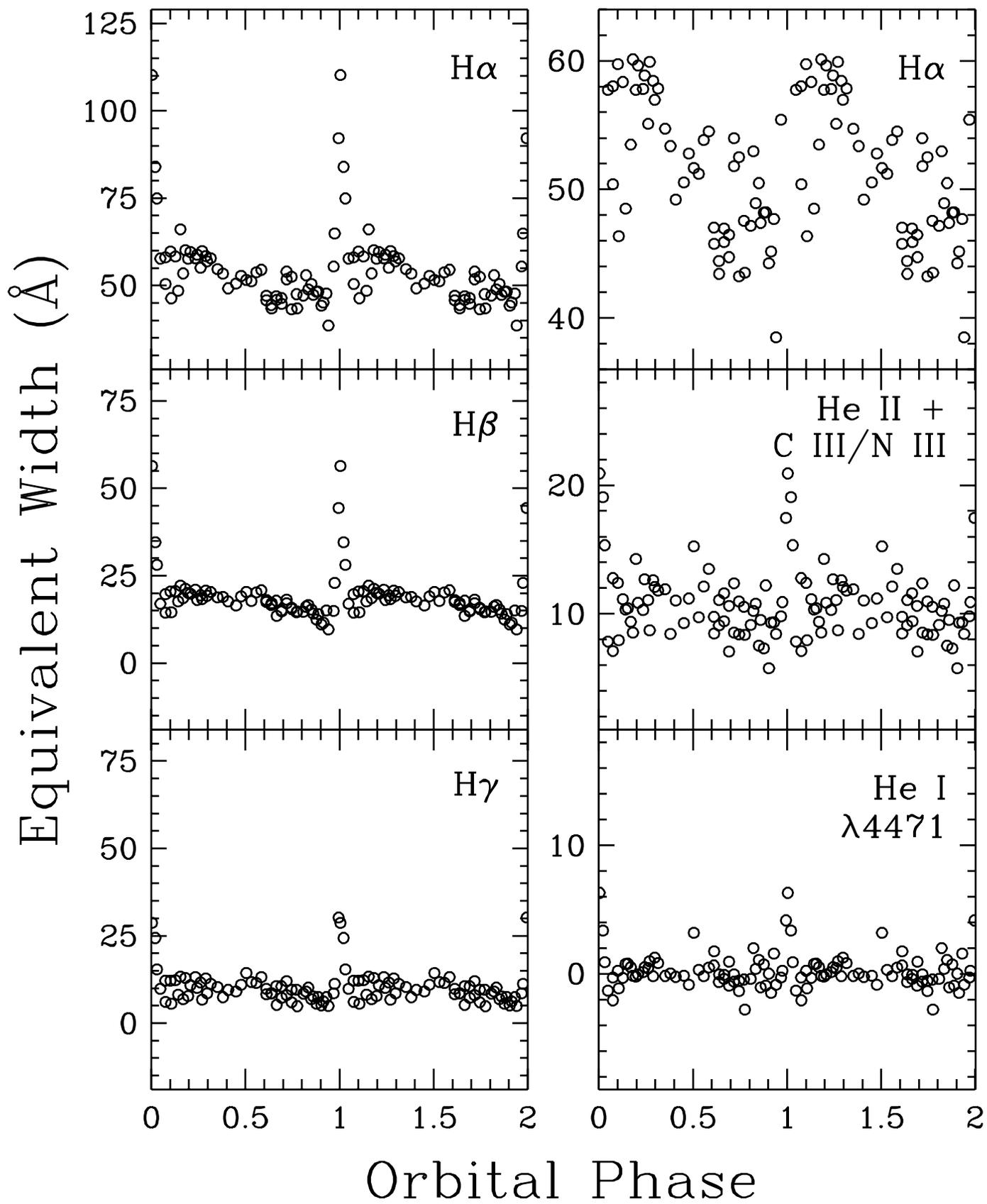
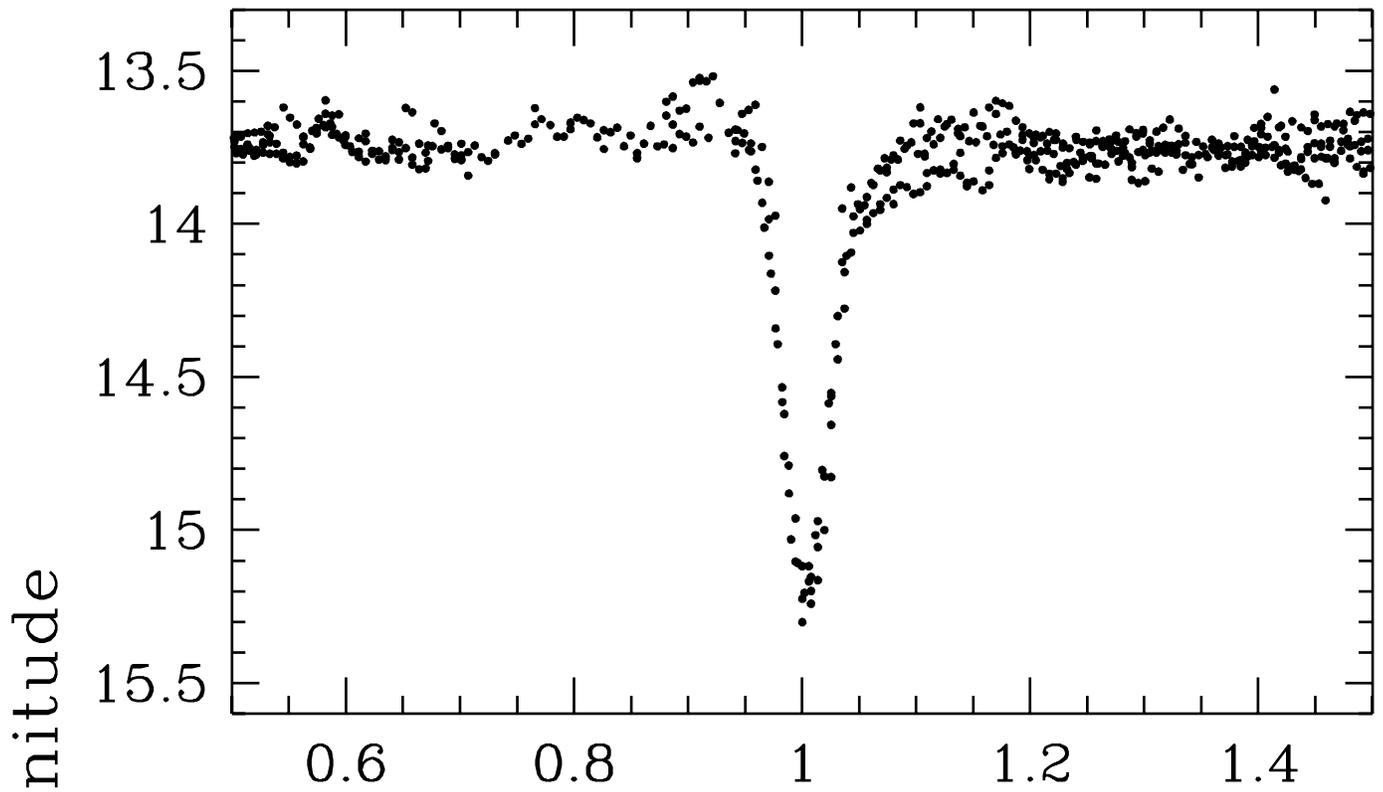
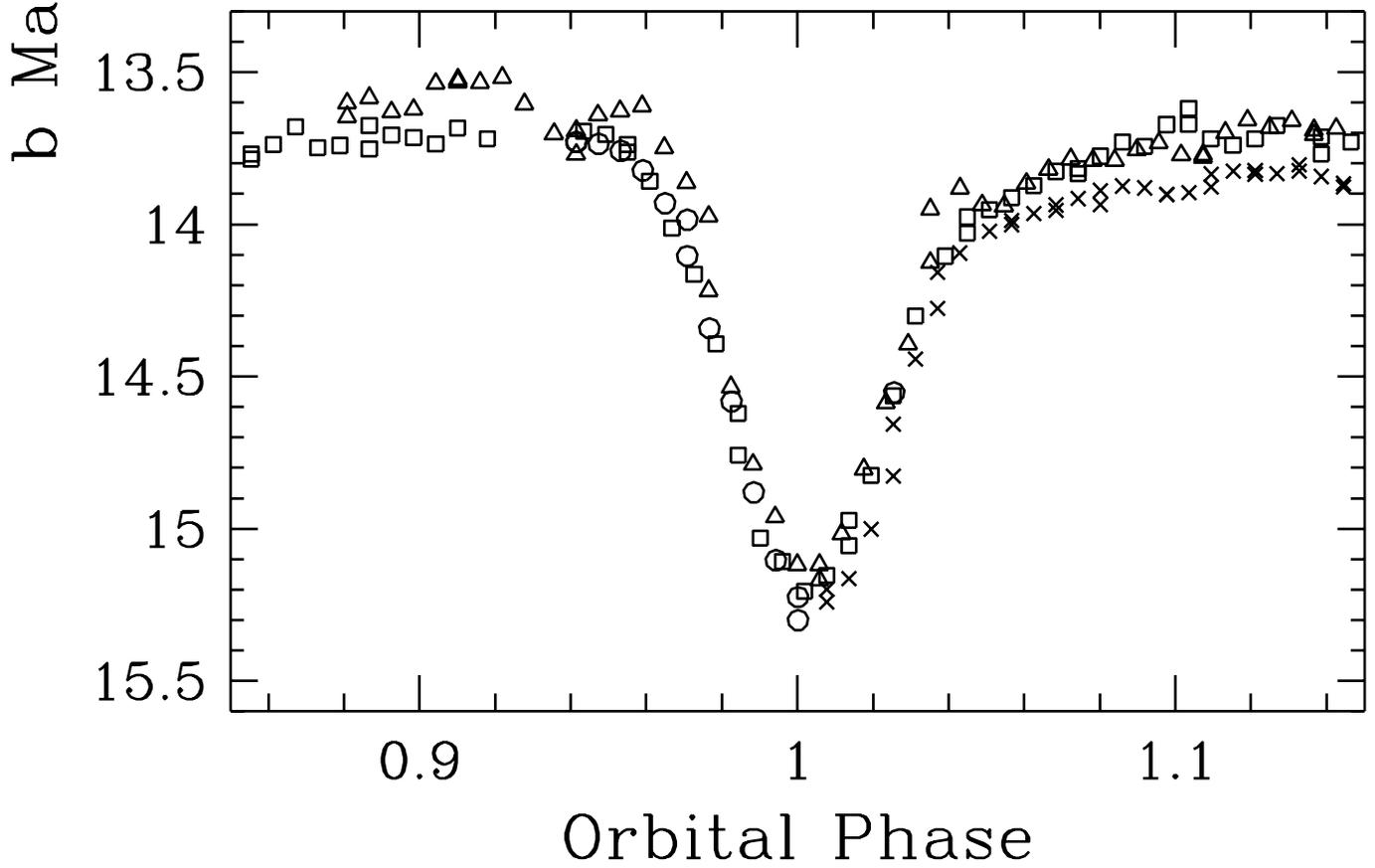
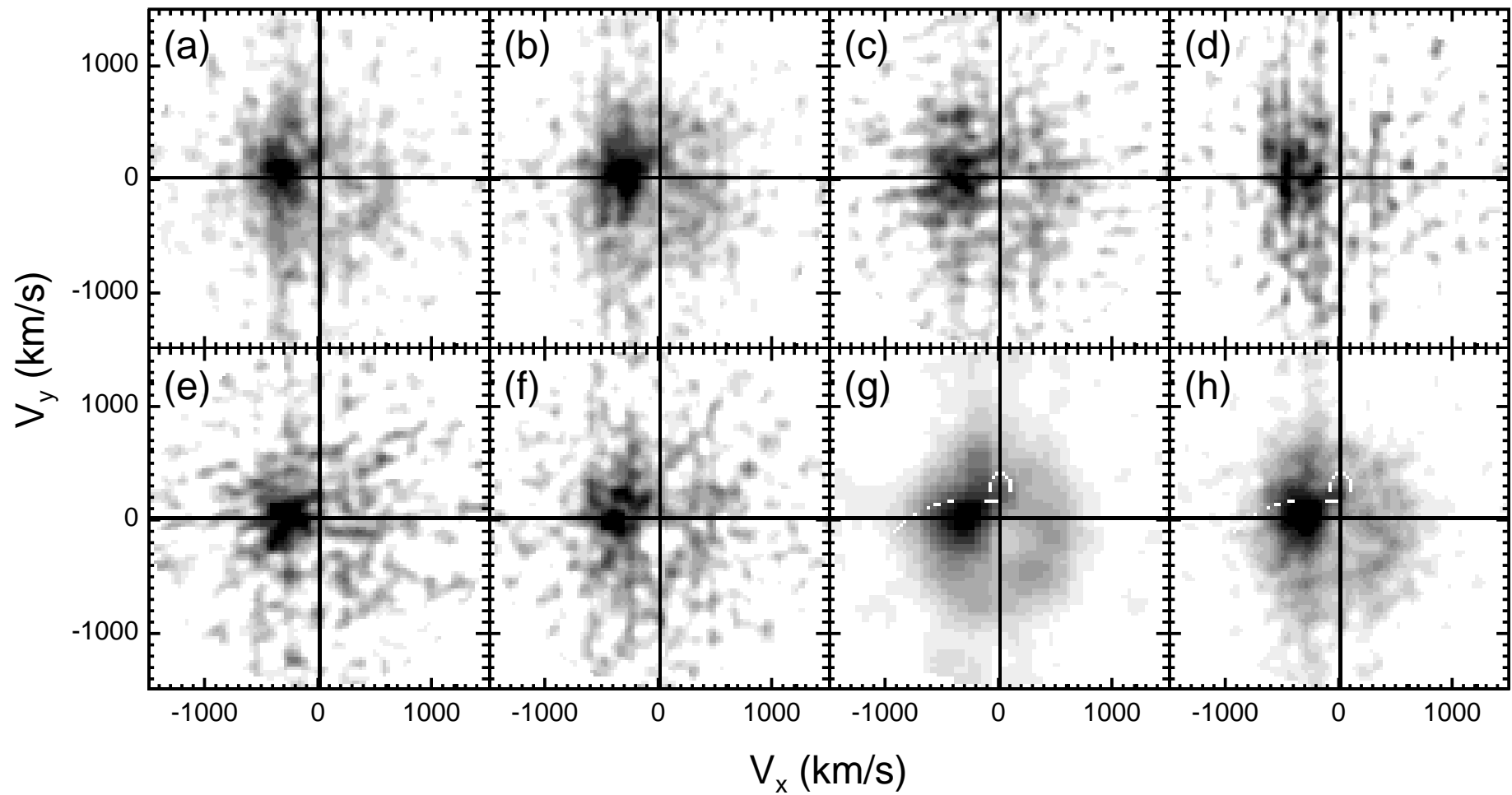
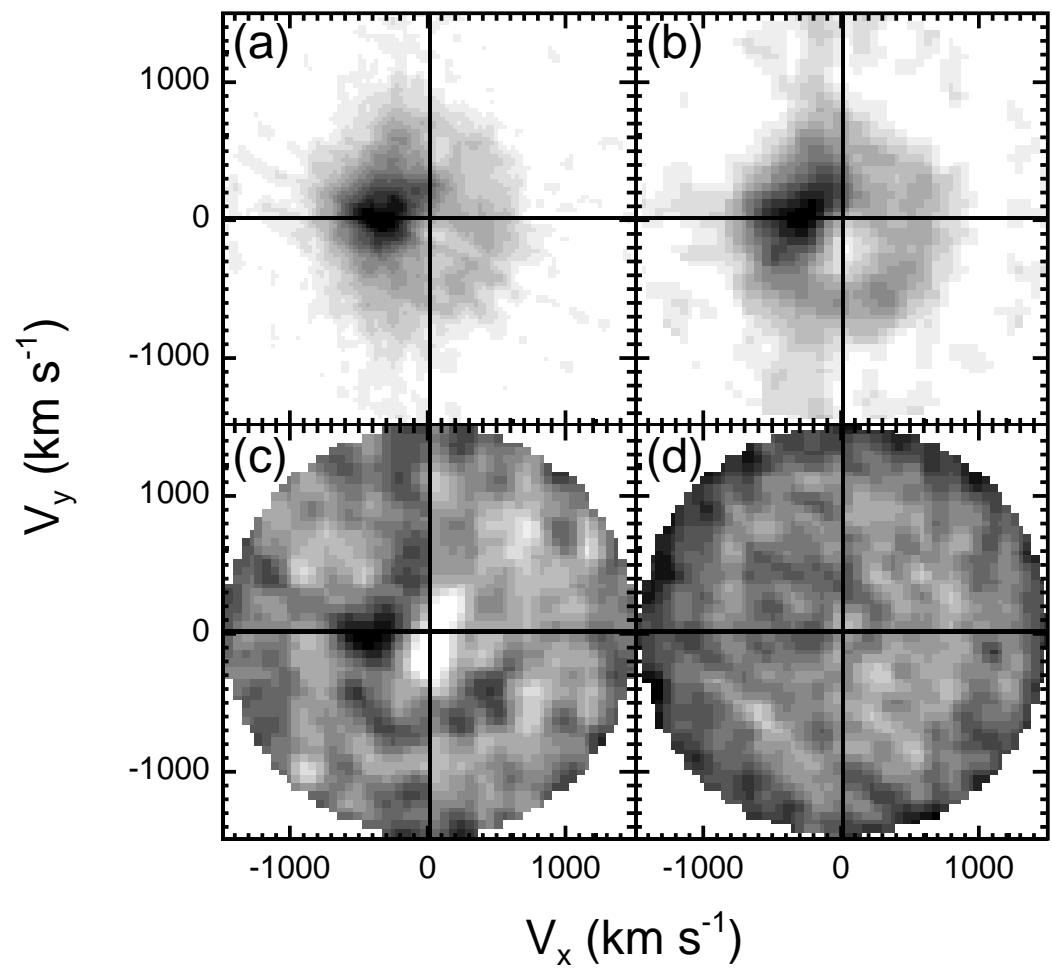
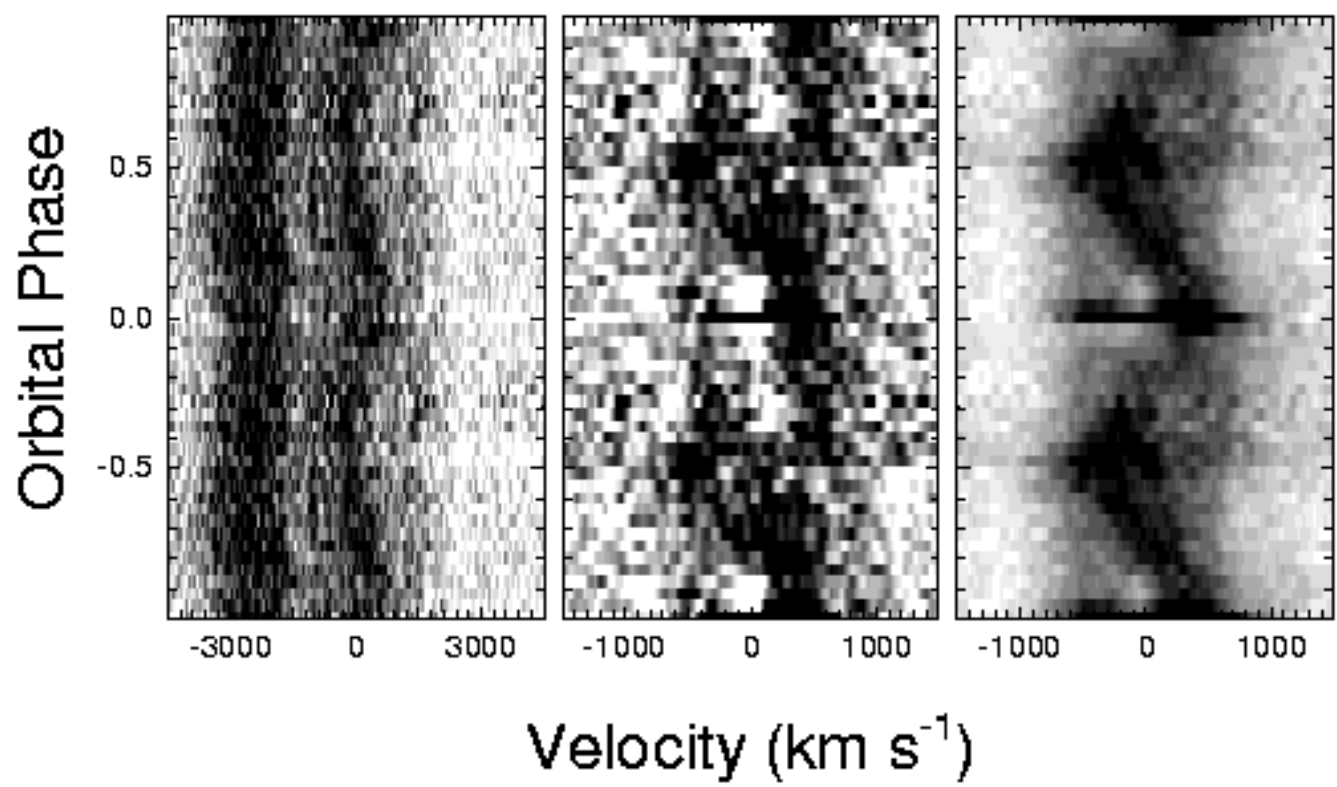
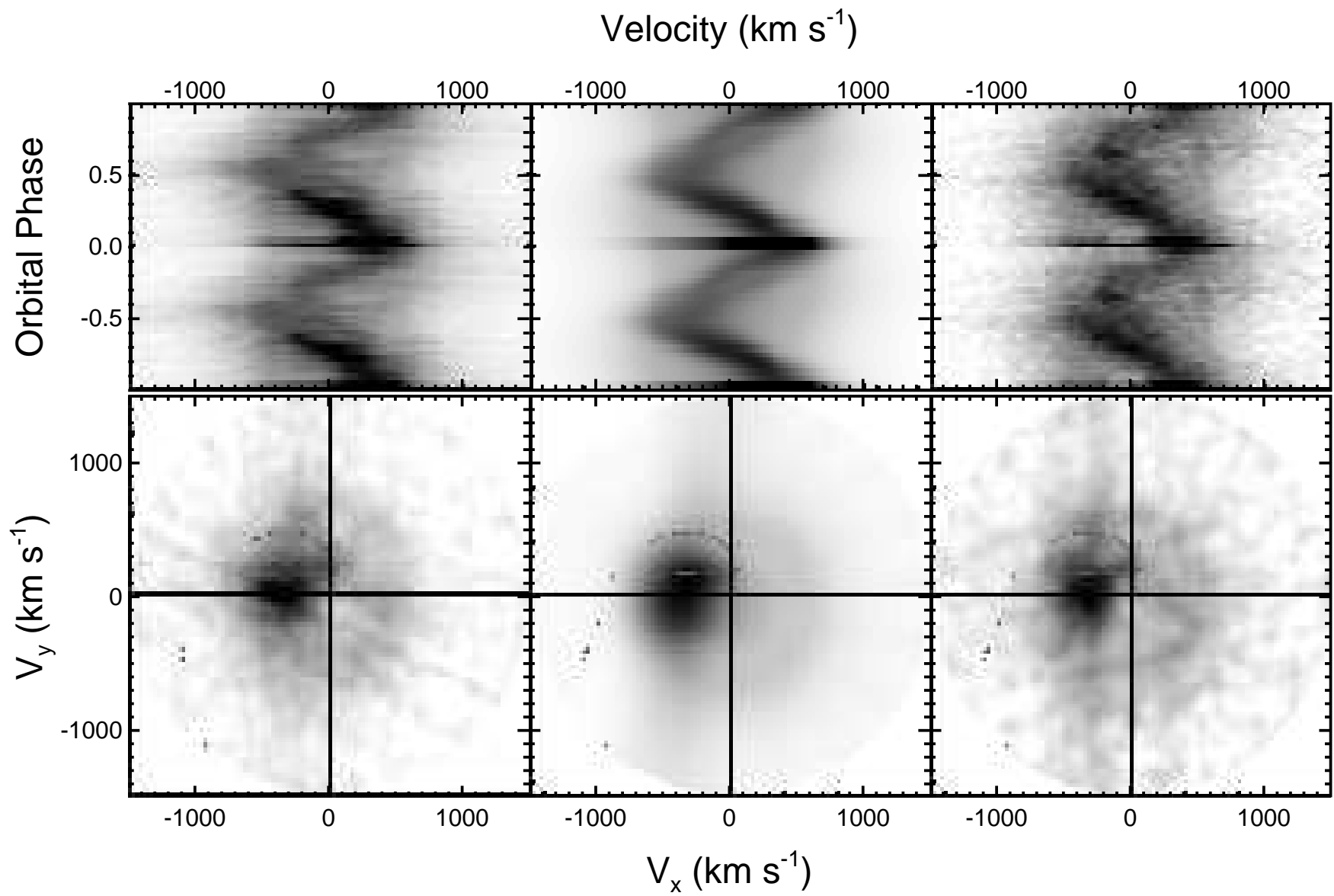
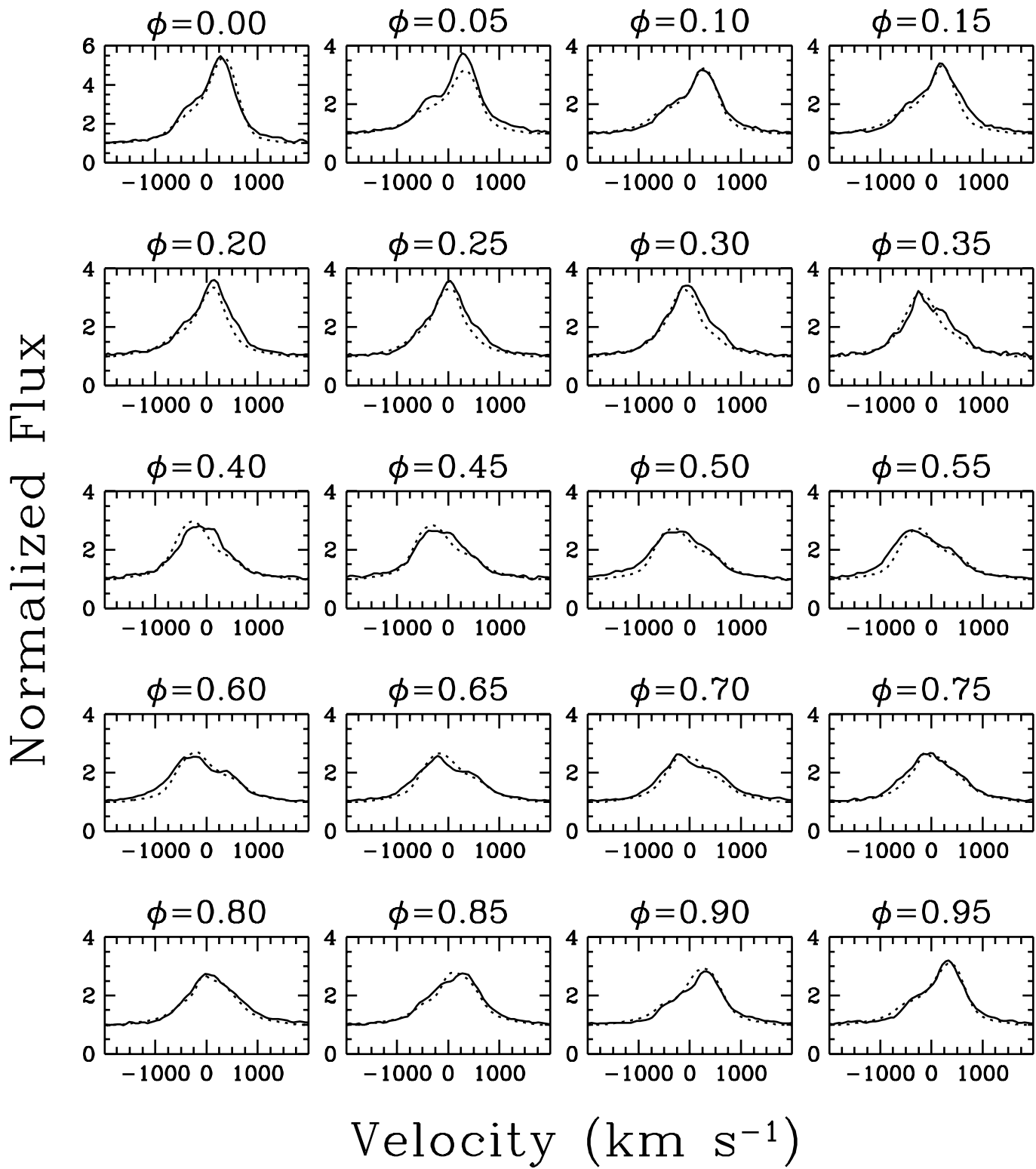
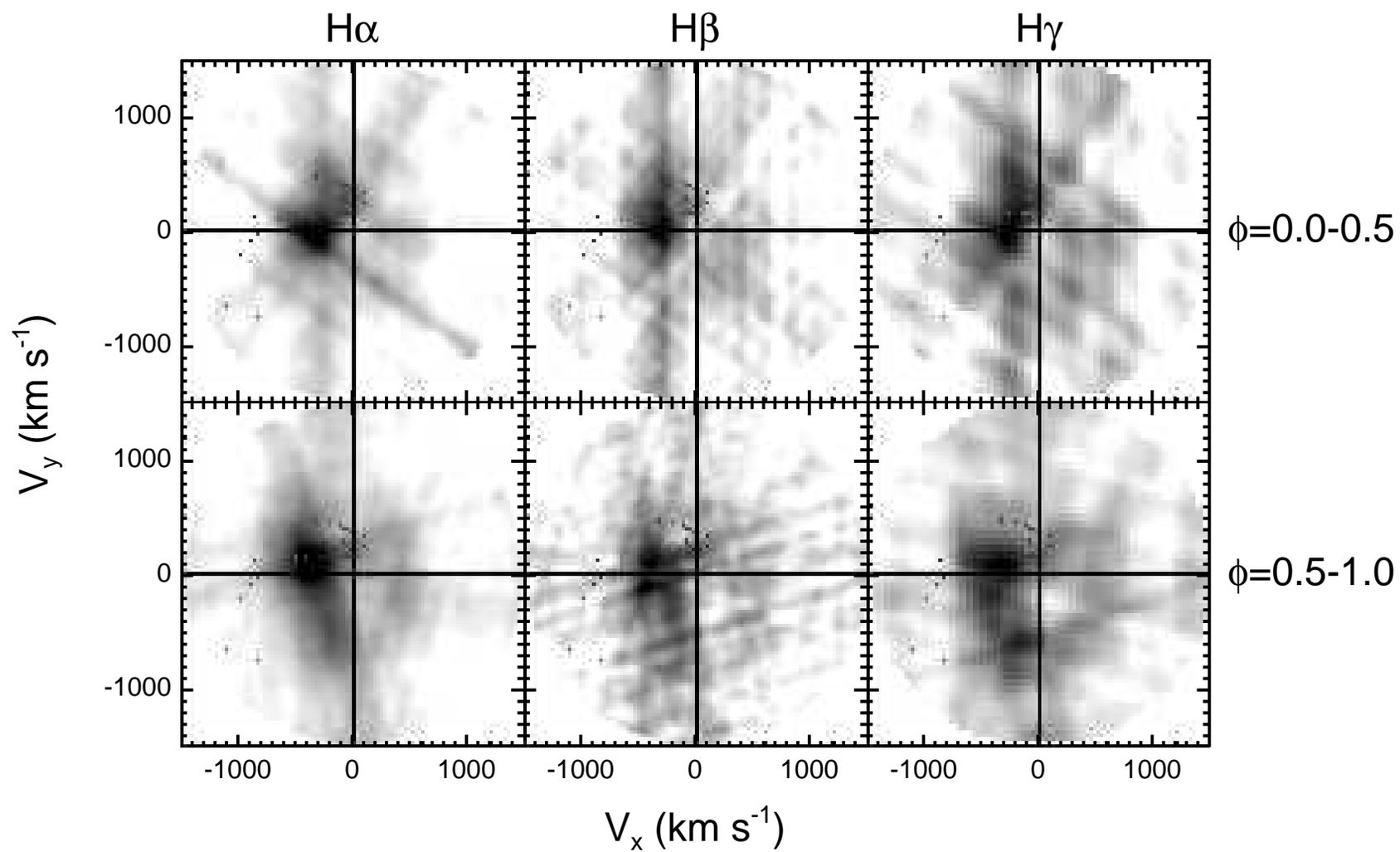
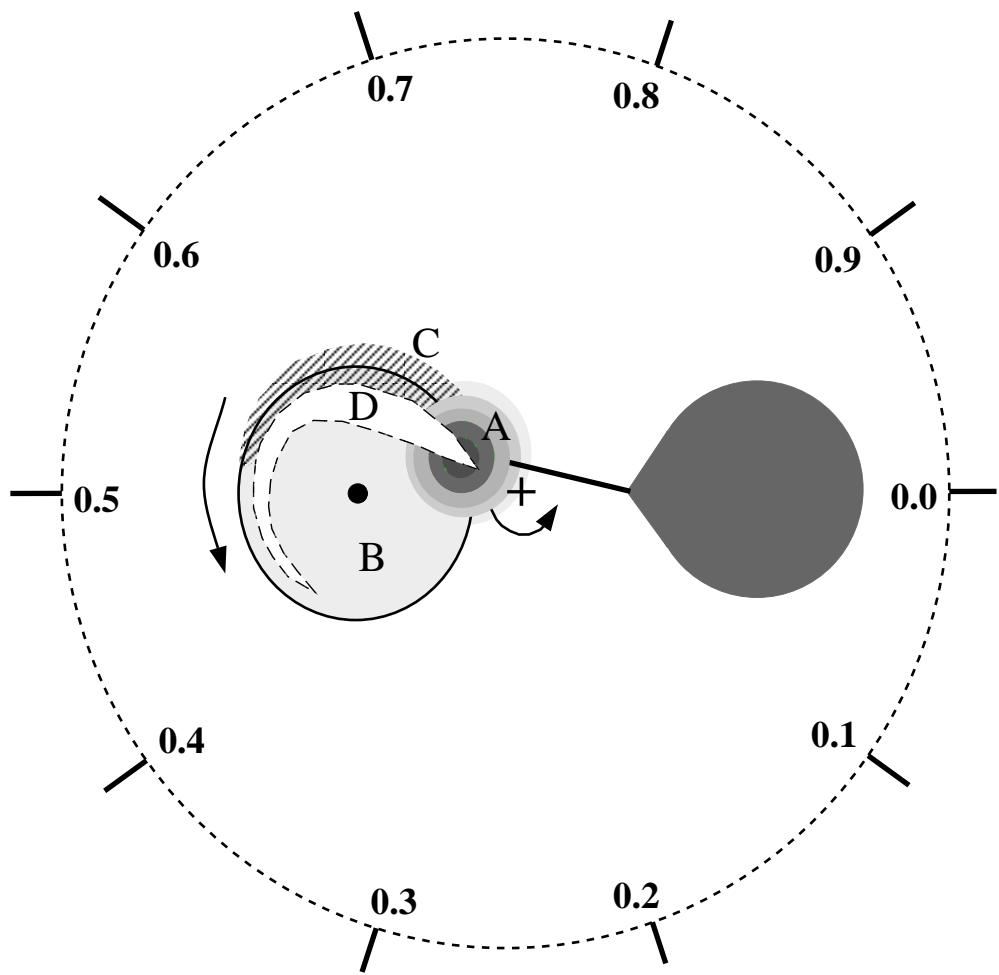
\end{document}